\documentclass[trackchanges, twocolumn, twocolappendix]{aastex701}

\usepackage{longtable, array, hyperref}
\begin{document}

\title{A Featureless JWST NIRISS/SOSS Spectrum Reveals No Primary Atmosphere on the Hot Super-Earth TOI-1685b and Suggests Sequestration of Light Elements in the Interior}

\author[orcid=0009-0008-2624-0746,sname='Hanna Adamski']{Hanna Adamski}
\affiliation{Department of Physics \& Astronomy, The University of California Los Angeles, Los Angeles, CA 90095, USA}
\affiliation{Department of Earth, Planetary, and Space Sciences, The University of California Los Angeles, Los Angeles, CA 90095, USA}
\email[show]{hadamski@g.ucla.edu}  

\author[0000-0002-0298-8089]{Hilke E. Schlichting}
\affiliation{Department of Earth, Planetary, and Space Sciences, The University of California Los Angeles, Los Angeles, CA 90095, USA}
\email[noshow]{hilke@g.ucla.edu} 

\author[orcid=0000-0001-5578-1498]{Björn Benneke}
\affiliation{Department of Earth, Planetary, and Space Sciences, The University of California Los Angeles, Los Angeles, CA 90095, USA}
\email[noshow]{bbenneke@ucla.edu} 

\author[0000-0002-2195-735X]{Louis-Philippe Coulombe}
\affiliation{Institut Trottier de recherche sur les exoplanètes (IREx), Université de Montréal, 1375 Ave Thérèse-Lavoie-Roux, Montréal, QC, H2V 0B3, Canada}
\email[noshow]{louis-philippe.coulombe@umontreal.ca} 

\author[0009-0002-2380-6683]{Eshan Raul}
\affiliation{Department of Astronomy, University of Wisconsin--Madison, Madison, WI 53706, USA}
\email[noshow]{eraul@umich.edu} 

\author[0000-0002-3328-1203]{Michael Radica}
\altaffiliation{NSERC Postdoctoral Fellow}
\affiliation{Department of Astronomy \& Astrophysics, University of Chicago, 5640 South Ellis Avenue, Chicago, IL 60637, USA}
\email[noshow]{radicamc@uchicago.edu} 

\author[0000-0002-2875-917X]{Caroline Piaulet-Ghorayeb}
\altaffiliation{E. Margaret Burbidge Postdoctoral Fellow}
\affiliation{Department of Astronomy \& Astrophysics, University of Chicago, 5640 South Ellis Avenue, Chicago, IL 60637, USA}
\email[noshow]{carolinepiaulet@uchicago.edu} 

\author[0000-0001-6809-3520]{Pierre-Alexis Roy}
\affiliation{Department of Earth, Planetary, and Space Sciences, The University of California Los Angeles, Los Angeles, CA 90095, USA}
\email[noshow]{paroy@g.ucla.edu}  

\author[orcid=0000-0002-9258-5311]{Thaddeus D. Komacek}
\affiliation{Department of Physics (Atmospheric, Oceanic and Planetary Physics), University of Oxford, Parks Rd, Oxford, OX1 3PU, UK}
\email[noshow]{tad.komacek@physics.ox.ac.uk} 

\author[orcid=0000-0001-9291-5555]{Charles Cadieux}
\affiliation{Institut Trottier de recherche sur les exoplanètes (IREx), Université de Montréal, 1375 Ave Thérèse-Lavoie-Roux, Montréal, QC, H2V 0B3, Canada}
\affiliation{Observatoire de Gen\`eve, D\'epartement d’Astronomie, Universit\'e de Gen\`eve, Chemin Pegasi 51, 1290 Versoix, Switzerland}
\email[noshow]{charles.cadieux.1@umontreal.ca}

\author[orcid=0000-0003-0973-8426]{Eva-Maria Ahrer}
\affiliation{Max-Planck-Institut f\"{u}r Astronomie, K\"{o}nigstuhl 17, 69117 Heidelberg, Germany}
\email[noshow]{ahrer@mpia.de}

\author[0000-0001-5442-1300]{Thomas M. Evans-Soma}
\affiliation{School of Science, University of Newcastle, Callaghan, NSW, Australia}
\email[noshow]{Tom.Evans-Soma@newcastle.edu.au}

\author[orcid=0000-0002-0747-8862]{Yamila Miguel}
\affiliation{Leiden Observatory, University of Leiden, Einsteinweg 55, 2333CA Leiden, The Netherlands}
\affiliation{SRON Netherlands Institute for Space Research , Niels Bohrweg 4, 2333CA Leiden, the Netherlands}
\email[noshow]{ymiguel@strw.leidenuniv.nl} 

\author[orcid=0000-0001-5383-9393]{Ryan Cloutier}
\affiliation{Department of Physics \& Astronomy, McMaster University, 1280 Main St West, Hamilton, ON, L8S 4L8, Canada}
\email[noshow]{ryan.cloutier@mcmaster.ca}

%\author{JWST Water World}

%%% 250 Word Limit
\begin{abstract}
The frequency and properties of atmospheres around close-in rocky planets orbiting late-type stars remain poorly understood. Transmission spectroscopy provides a means of searching for such atmospheres by identifying atomic and molecular absorption features. Here, we present the first JWST NIRISS/SOSS (0.6—2.8$\mu$m) transmission spectrum of the hot (T$_{\mathrm{eq}} \sim 1060$ K), short-period super-Earth TOI-1685b (M$_{\mathrm{p}} = 3.03\pm0.33 \mathrm{M}_\oplus$, R$_\mathrm{p} = 1.43\pm0.04 \mathrm{R}_\oplus$), orbiting an M2.5V star. Despite uncertainties of 20 ppm for 30-nm bins, we detect no spectral features beyond a broad trend attributable to stellar contamination. Combined with previous JWST/NIRSpec transit observations, these results are compatible with the absence of a substantial atmosphere, high-altitude gray clouds, or a compact atmosphere with mean molecular mass above 7 g/mol ($> 3\sigma$).  Chemical-equilibrium retrievals rule out cloud-free, low-mean-molecular-mass atmospheres with surface pressures above 1 mbar. The planet’s bulk density, scaled to $1 \mathrm{M}_\oplus$, is $84\pm14 \%$ of Earth’s ($\rho_{1M_\oplus} = 4.65 \pm 0.82 \ \mathrm{g/cm^3}$), indicating a density deficit relative to an Earth-like composition. In the absence of a volatile envelope, this deficit can be explained by either a reduced iron core mass fraction (CMF $\leq 0.22$) or light element sequestration into the interior, corresponding to 1.3—3 wt$\%$H and 6.2—14 wt\%O in the metal core. We favor the latter scenario because the host star’s high metallicity disfavors low CMFs. TOI-1685b thus joins the growing population of close-in rocky planets around M-dwarfs with no detectable atmospheres.

\end{abstract}



\section{Introduction} 
The occurrence rate of small exoplanets exhibits a bimodal distribution, consisting of super-Earths and sub-Neptunes separated by the ``radius valley," a dearth of planets with radii roughly between 1.8 and 2.0 $R_\oplus$ first identified for planets orbiting FGK-type stars \citep{radiusgap, 2018fultonpetig}. One interpretation suggests that both populations originated from a single group of planets with extended, H$_2$-dominated envelopes, with the valley produced by atmospheric mass-loss processes---either via photoevaporation, driven by high-energy stellar irradiation \citep{2017owen}, or through core-powered mass loss, where the residual heat from formation together with the stellar bolometric luminosity erodes planetary envelopes \citep{2018ginz, 2020gup}. These mechanisms may act synergistically, with planets transitioning between the two regimes during their evolution \citep{2024owen}. Alternatively, the valley may reflect distinct formation pathways \citep[e.g.][]{2018fultonpetig, 2019gaspoor}. Around lower-mass stars, particularly M-dwarfs, the valley appears more populated and may shift in location, with several planets residing within the nominal gap having bulk densities consistent with substantial volatile inventories in the envelope and/or interior \citep{2022luque, 2023rogers, ho2024, 2024venturini, 2024parc}. 

Despite the emerging diversity of planets around M-dwarfs, it remains unclear whether rocky planets in these environments have observable atmospheres. To date, transmission spectroscopy has not yielded definitive detections of molecular absorption features on small, close-in planets \citep{2023lustig, Alderson_2024, Cadieux_2024, Radica_1c_2025, Piaulet-Ghorayeb_2025, 2025allen}, while thermal emission measurements often report albedos and day-night temperature gradients consistent with bare-rock surfaces \citep{2019kreidberg,Crossfield_2022, 2023Natur.618...39G, 2025emi}. This lack of detections may not be surprising, as many rocky M-dwarf planets reside on ultra-short-period (USP) orbits---with $P<1$ day $\&$ $R_p \leq 2R_\oplus$---where the incident flux can exceed that received by Earth by several orders of magnitude \citep{2020gronoff, hanson25}.
%Additionally, the very presence of USPs at small orbital separations also challenges formation theory: in-situ formation is unlikely given their location inside the dust sublimation line, making migration scenarios---such as planet-planet scattering, Kozai-Lidov cycles, or combinations thereof---the most plausible formation pathways \citep{2021usp, Zhu_2025, 2025songhu}. Determining the compositions of surviving atmospheres may therefore offer insights into their migration histories and formation environments. 

If an atmosphere is detected on a highly irradiated rocky planet, its composition can place strong constraints on both the origin and evolution of its envelope. For a population that initially accreted H$_2$-dominated atmospheres, extreme irradiation drives hydrodynamic escape that preferentially removes hydrogen, leading to mass-dependent fractionation of the remaining material and progressive enrichment in heavier species such as deuterium and helium, with the greatest enrichment expected for planets retaining only tenuous envelopes \citep{lrogers_frac, 2023_mass_frac_he, 2024_cherubim_massfrac_he_d}. Sustained hydrogen loss also perturbs the global redox balance, coupling atmospheric escape to interior chemical evolution and driving progressive oxidation. Through this coupled evolution, \citet{fleeting_not_forgotten} show that as a sub-Neptune loses its envelope to become a super-Earth the residual atmosphere is expected to become dominated by H$_2$O by mass. 

Alternatively, for planets that have completely lost their primordial envelopes, secondary atmospheres may arise via volcanic outgassing, impact delivery of volatiles \citep{2018kral, 2018schlichting}, or radiative cooling from a molten surface \citep{2024chat}. These mechanisms provide pathways for atmospheric regeneration, even after significant erosion, though the longevity of such secondary atmospheres remains uncertain under the intense stellar activity typical of M-dwarf systems \citep{2023lim}.  

With JWST, the sensitivity and spectral coverage now exist to test whether the most favorable rocky planets orbiting M-dwarfs host detectable atmospheres. Although their small, cool host stars provide favorable planet-to-star radius ratios, observations to date largely favor bare rock scenarios, with only tentative evidence for atmospheres around a handful of short-period rocky worlds. For example, the Hot Rocks program (JWST GO 3730; PI Diamond-Lowe) used MIRI’s 15 micron filter to observe the secondary eclipses of nine irradiated terrestrial planets orbiting M-dwarfs; three are consistent with bare rock surfaces \citep{2025bare_rock, 2025fortune,2025allen}, while only one shows plausible evidence for an atmosphere \citep{2025August}. Further characterization of low-density, short-period rocky planets such as TOI-1685b is crucial for constraining atmospheric retention efficiencies, evaluating the role of interior-atmosphere coupling in generating atmospheric observables, and quantifying the diversity of super-Earths in extreme close-in orbits.

\subsection{The TOI-1685 System}
TOI-1685b is a hot ($T_\mathrm{eq} \approx 1062$\,K; \citealt{Burt_2024}), tidally locked super-Earth transiting an early-type M2.5V star.  It is the only confirmed planet in this system, with an orbital period of approximately $P_\mathrm{orb} \approx 0.669$ days \citep{bluhm_2021, Hirano_2021, Burt_2024}. The host star is bright, nearby, and relatively inactive, with a $J_\mathrm{mag} = 9.616 \pm 0.022$ (2MASS All-Sky Survey; \citet{2003cutri}) and a distance of $37.596 \pm 0.022$ pc \citep{2021bailer}. Stellar rotation measurements yield $P_\mathrm{rot} = 18.2 \pm 0.5$ days \citep{Burt_2024}, indicating low magnetic activity. The observed rotational period is further supported by the star's slow projected rotational velocity, ($v\sin i < 2$\,km/s; \citealp{2021marfil, bluhm_2021}). 

Since its discovery by the Transiting Exoplanet Survey Satellite (TESS; Sector 19), TOI-1685b has been characterized by discrepant estimates of its radius and bulk density \citep{bluhm_2021, Hirano_2021}. Using TESS photometry and CARMENES radial velocities, \citet{bluhm_2021} measured a radius of $R_P = 1.70 \pm 0.07 R_\oplus$ and a bulk density of $\rho_P = 4.21^{+0.95}_{-0.82}$ g/cm$^3$, consistent with a $50\% \ $H$_2$O / $50\% \ $ MgSiO$_3$ composition. In contrast, \citet{Hirano_2021} found a smaller radius of $R_P = 1.459 \pm 0.065 R_\oplus$ and a higher density of $\rho_P = 6.1^{+1.9}_{-1.7} \ $g/cm$^3$, suggesting a largely rocky composition similar to Earth's. A joint analysis of the available transit and radial-velocity data by \citet{2022luque}, favored the larger-radius, lower-density interpretation, further supporting the presence of a volatile component in TOI-1685b's interior.

Recent measurements of the metallicity of TOI-1685 present a somewhat conflicting picture of its host star, with \citet{Burt_2024} reporting a metal-rich composition of [Fe/H] $= 0.3 \pm 0.1$, while more recent measurements by \citet{gromek2026} favor a near-solar metallicity of [M/H] $= 0.06 \pm 0.08$. Using new observations with the MAROON-X spectrograph \citet{Burt_2024} also revised the stellar parameters, deriving a mass of $0.454 \pm 0.018 \ M_\odot$ and a radius of $0.4555 \pm 0.0128 \ R_\odot$. The revised stellar parameters yield smaller estimates for TOI-1685b, with a radius of $1.468 \pm 0.05\,R_\oplus$ and mass of $3.03^{+0.33}_{-0.32}\,M_\oplus$, corresponding to a bulk density of $\rho_P = 5.3 \pm 0.8$\,g/cm$^3$ ($0.96 \rho_\oplus$; $\rho_\oplus = 5.51$\,g/cm$^3$) \citep{Burt_2024}. Independent CHEOPS observations \citep{2025egger} yield a radius of $1.421 \pm 0.06\,R_\oplus$, consistent with previous measurements by \citet{Hirano_2021} and \citet{Burt_2024}. Together, these measurements place TOI-1685b squarely in the category of hot, rocky super-Earths with negligible volatile envelopes, rather than water-rich worlds. 

% Its atmospheric characterization can therefore constrain whether highly irradiated rocky planets retain secondary atmospheres and, if so, how atmospheric escape and interior-atmosphere coupling shape their observable properties.
Taken together, these measurements make TOI-1685b a compelling target for spectroscopic characterization. Its close-in orbit exposes it to extreme stellar irradiation, while its location near the radius valley provides an opportunity to test atmospheric survival across the super-Earth/sub-Neptune transition. The planet has been observed in three JWST Cycle 2 programs, including a full NIRSpec/G395H phase curve (GO 3263, PI: Luque), four NIRSpec/G395H transits (GO 4195, PI: Fisher), and two NIRISS/SOSS transits (GO 4098, PI: Benneke). 

From the phase curve observations, \citet{luque} find negligible heat redistribution and a low albedo, with the observed dayside brightness temperature consistent with blackbody-like emission from an airless, rocky surface. The NIRSpec/G395H transmission analysis of \citet{2026fisher} likewise finds a largely featureless spectrum and confidently rules out hydrogen-dominated atmospheres, while cloudy or high-mean-molecular-weight secondary atmospheres remain viable. Thus, while current observations favor a barren, rocky surface, they do not uniquely exclude a secondary atmosphere, highlighting the need for additional spectroscopic constraints on atmospheric composition and structure. In this study, we present the first NIRISS/SOSS transmission spectrum of TOI-1685b, along with an updated analysis of its atmospheric properties.

This paper is structured as follows: In Section \ref{sec:obs_calibs}, we describe the JWST NIRISS/SOSS observations and detail the data reduction procedures using two independent pipelines. We also re-analyze the published JWST NIRSpec/G395H dataset from \citet{luque}, processing it with our selected pipelines and outlining modifications to the default reduction steps. In Section \ref{sec:exotep}, we present our white-light and spectroscopic light curve fitting methodologies for both the NIRISS and NIRSpec datasets. Atmospheric retrieval techniques and analyses are discussed in Section \ref{sec:atmsec}, with results summarized in Section \ref{sec:results}. In Section \ref{sec:mr_relations}, we examine the most plausible atmospheric and interior compositions of TOI-1685b based on our findings, with the main conclusions summarized in Section \ref{sec:conclusion}.

\section{Observations and Data Calibration} \label{sec:obs_calibs}
In this work, we analyze two NIRISS/SOSS transit observations of TOI-1685b using two independent reduction pipelines---\texttt{exoTEDRF} and \texttt{NAMELESS} \citep[as in ][]{Piaulet-Ghorayeb_2024, Ahrer_2025}. A detailed discussion of the \texttt{NAMELESS} reduction is presented in Appendix \ref{appendix:NAMELESS_NIRISS} (Figure \ref{appendix:NL_EX_comp}). We additionally incorporate a NIRSpec transit observation from a previous Cycle 2 JWST Program (GO-3263; \citet{luque}) into our analysis. Our atmospheric retrievals use the \texttt{exoTEDRF} transmission spectra as our baseline dataset. To assess the robustness of our results to the data reduction procedure, we also compare the transmission spectra produced by the two pipelines and verify that our conclusions are not sensitive to the choice of reduction pipeline.  

\subsection{NIRISS/SOSS} \label{sec:exotedrf_niriss}
Two transits of TOI-1685b were observed using the Single Object Slitless Spectroscopy Mode (SOSS) of NIRISS \citep{Doyon_2023, Albert_2023} as a part of the Cycle 2 General Observer program GO 4098 (PI: Benneke, co-PI: Evans-Soma). Each NIRISS/SOSS time-series observation (TSO) consists of 320 integrations, each comprising 5 groups, and spans approximately 2.9 hr. The choice of the SUBSTRIP256 array facilitates the extraction of the stellar spectrum across both order 1 (0.8-2.85 $\mu$m) and order 2 (0.6-1.4 $\mu$m) at medium resolving power (R $\approx$ 650 at 1.25 $\mu$m; \citealp{Albert_2023}). The visits occurred on UTC 2025 January 15 and 16, and covered 1.4 hr of baseline before the transit, 0.9 hr in transit, and 0.6 hr out of transit. The two transits were separated by 16 hours, corresponding to approximately $3.6\%$ of the stellar rotation period. This short interval between the observations was advantageous, as it reduced the likelihood of significant changes in stellar activity between the two datasets and between the extracted transmission spectra. 

Each SOSS TSO is reduced using the \texttt{exoTEDRF} pipeline\footnote{\url{https://exotedrf.readthedocs.io/en/latest/index.html}} \citep{2023ad, Radica_2023, Radica_2024}, closely following the procedure carried out in \citet{Piaulet-Ghorayeb_2024} and \citet{2024a}. We perform the default detector-level and spectroscopic-level calibrations captured by \texttt{Stage1} and \texttt{Stage2}, respectively, and outline any significant modifications here. Cosmic ray hits are removed using a time-domain sigma-clipping algorithm with a $7\sigma$ threshold \citep{Radica_2024}. Additionally, we perform background and 1/f noise corrections, with the piece-wise background rescaling yielding correction factors of 0.91368/0.80977 and 0.88717/0.81900 for Visits 1 and 2, respectively. A more robust, column-wise 1/f correction using the \texttt{scale-achromatic} method is implemented at the integration level. Furthermore, we interpolate flagged pixels by performing spatial and temporal outlier clipping (10$\sigma$ and 12$\sigma$, respectively). 

Principal component analysis (PCA) is performed on the calibrated two-dimensional time-series frames. The second and third principal components reveal variations in the y-position and FWHM of the trace, occurring near mid-transit in Visit 1 and at the start of Visit 2. Because these time-variable systematics cannot be reliably modeled or de-trended, we discard the affected integrations. For Visit 2, we remove the first 50 integrations from both Order 1 and Order 2 light curves, while for Visit 1, we mask integrations 215 to 230. The masked regions are shown in gray in Appendix Figure \ref{appendix:raw_integ}. The origin of the short-term systematic remains uncertain, but its transient nature and subsequent return to the pre-event baseline appear inconsistent with previously observed mirror-tilt artifacts \citep{2023alderson}, which typically produce persistent flux offsets. The stellar spectrum is extracted using a 30-pixel-wide box aperture, minimizing the scatter in the out-of-transit baseline. Although there is minor overlap between Order 1 and 2 trace positions, the degree of self-contamination is negligible \citep{darveau2022, radica2022}. Additionally, we confirm that no significant dispersed or undispersed contaminants of Orders 0 or 1 are present.

\subsection{NIRSpec/G395H} \label{sec:exotedrf_nirspec}
TOI-1685b was also observed with JWST NIRSpec \citep{birkman2022nirspec, jakobsen2022nirspec} as a part of the Cycle 2 Program GO-3263 (PI: Luque; \citealt{luque}). The Bright Object Time Series mode was used analogously with the G395H grating to obtain the planet's phase curve on UTC 2024 February 14-15 between 23:37 and 18:36. The observation consisted of 4450 integrations, made up of 16 groups, and spanning 18.97\,hr. We reduced the full phase curve of TOI-1685b following the default steps described in \citet{Radica2024a, Ahrer_2025}, however we focus on the region surrounding the transit and encourage the reader to consult \citet{luque} for additional details captured by the full phase-curve analysis and the emission spectrum. 

\texttt{Stage1} detector-level calibrations are performed across both detectors, omitting reference pixel calibrations as they introduce row-dependent noise. As in \citet{toi270b} for TSOs observed with NIRSpec, we correct for 1/f noise at the group-level, masking all flagged outliers and pixels within 7 pixels above or below the target spectral trace. We use a time-domain clipping threshold of $10\sigma$ to detect and correct each frame for cosmic ray hits. Minor modifications to the wavelength solution which correct for the target star's misalignment within the slit, are captured in \texttt{Stage2}. We repeat the 1/f noise and background corrections at the integration level. The resulting noise properties are only minimally improved after this second correction, indicating that our group-level 1/f noise removal was sufficient. We interpolate pixel outliers using a $10\sigma$ spatial and temporal threshold criteria, consistent with previous NIRSpec reductions \citep{Radica_2024, Ahrer_2025, toi270b}.

We additionally apply PCA to NIRSpec/G395H data, finding negligible y-pixel drift in both detectors across their respective target traces over the duration of the TSO, consistent with a stable exposure. Finally, using the \texttt{edgetrigger} algorithm \citep{radica2022}, we locate the centroids of the spectral trace on NRS1 and NRS2, followed by the extraction of the stellar spectra using a simple box aperture with an extraction width of 10 pixels surrounding the target trace.

\section{\texttt{ExoTEP} Light Curve Analysis} \label{sec:exotep}
\label{sec:lcs}
We fit the transit light-curves of TOI-1685b extracted from the NIRISS and NIRSpec observations described above. For our light curve analysis, we focus exclusively on the transit portion of the phase-curve data and do not model the full phase variation. Specifically, we extract the integrations spanning the transit of TOI-1685b in the NIRSpec phase curve, from integration 2100 to 2700, corresponding to a time range of MJD$_\mathrm{TDB}$ 60355.35767-60355.46411. To better constrain periodic variability in the time-series data---likely due to stellar activity---we adopt a baseline approximately twice the duration of the transit. The transit observation spans 2 hours and 33 minutes, including 46 minutes of pre-transit and 56 minutes of post-transit baseline, centered on a $\sim51$ minute transit. We perform both white-light and spectroscopic light curve fits to the NIRISS/SOSS and archival NIRSpec/G395H datasets reduced using the \texttt{exoTEDRF} and \texttt{NAMELESS} pipelines as detailed in the preceding sections. 
 
%%%%%%%%%%%%%%%%%%%%%%%%%%%%%%%%%%%% Star + Planet
%%% Table of TOI-1685/b parameters
\begin{table}[]
\centering
\begin{tabular}{l c} 
 \hline \hline
Parameter & Value\\ [0.5ex] 
 \hline \hline
TOI-1685 (star) & \\
 \hline
 Spectral Type$^a$ & M2.5V \\ 
 Effective Temperature$^a$ $T_\mathrm{eff}$ (K) & 3575 $\pm$ 53\\
 Metallicity$^a$ [Fe/H] & 0.3 $\pm$ 0.1\\ 
 Stellar Mass$^a$ ($M_{\odot}$) & 0.454 $\pm$ 0.018\\ 
 Stellar Radius$^a$ ($R_{\odot}$) & 0.4555 $\pm$ 0.013\\ 
 $\log_{10}$($g_\star$[cm/s$^2$])$^b$ & 4.85 $\pm$ 0.04 \\ 
 Stellar Rotation Period$^b$ (days) & $18.66^{+0.71}_{-0.56}$\\ [1ex] 
 \hline
 TOI-1685 b (planet) &\\
 \hline 
 Orbital Period$^a$ $P$ (days) & 0.669 \\ 
 Scaled Semimajor Axis $a/R_\mathrm{s}$ & $5.39^{+0.20}_{-0.36}$\\
 Impact parameter $b$ & $0.35^{+0.10}_{-0.11}$\\
 Inclination $i$ (deg) & $87.69^{+1.65}_{-2.02}$\\
 Planetary Radius $R_\mathrm{p}$ ($R_\oplus$) & 1.427 $\pm$ 0.043 \\
 Planetary Mass$^a$ $M_\mathrm{p}$ ($M_\oplus$) & $3.03^{+0.33}_{-0.32}$ \\
 Equilibrium Temperature $T_{\mathrm{eq}, A_\mathrm{B} = 0}$ (K) & $1089^{+26}_{-40}$ \\
\hline 

\end{tabular}
\caption{Stellar and Planetary Parameters Derived or Adopted for This Work. (a) Parameters adopted from \citet{Burt_2024}. (b) Parameters adopted from \citet{bluhm_2021}. }
\label{tab:toi_params}
\end{table}

\begin{figure*}[t]
    \begin{center}
        \includegraphics[width=0.75\textwidth]{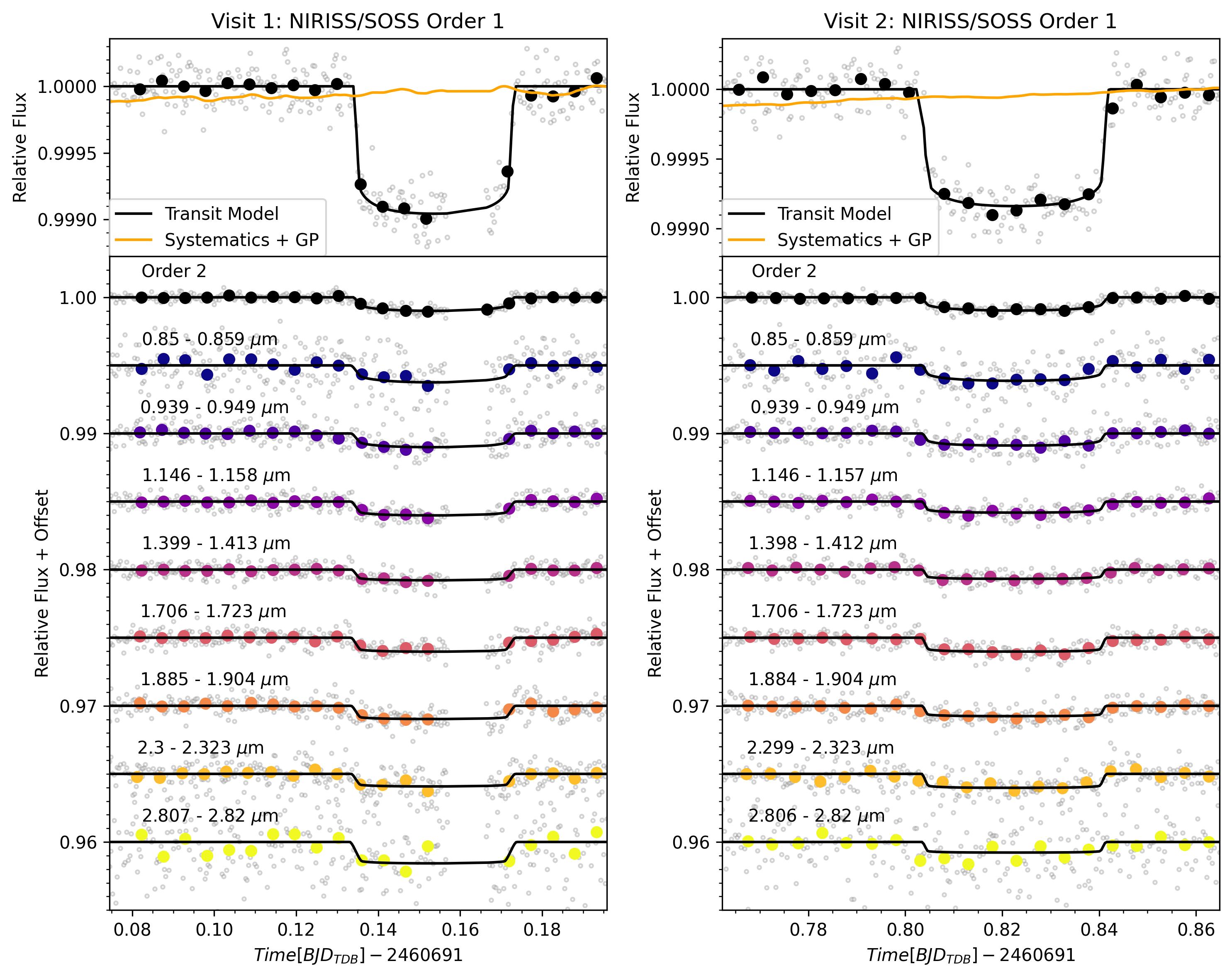}
        \caption{Broadband and spectroscopic light-curve fits of two NIRISS/SOSS transits of TOI-1685b from \texttt{exoTEDRF} reduction. \textit{Upper panels:} The systematics-corrected Order 1 broadband white-light curves are shown, binned down for visualization purposes (black), along with the best-fit astrophysical transit model (black) and the systematics model including the GP component (orange). \textit{Lower panels:} Results of eight systematics-corrected spectroscopic light-curve fits across Order 1 and the detrended white light curve for Order 2. The spectroscopic light curves are plotted with a relative flux offset and are binned temporally in groups of 20 points for visual clarity. The extent of each wavelength bin is defined above the best-fit model.}
        \label{fig:lightcurves}
    \end{center}
\end{figure*}

\subsection{Broadband Light Curve Fitting} \label{subsec:wlc}
Our primary goal is to measure the transmission spectrum of TOI-1685b. To achieve this, we perform identical joint fits of the light curves produced by \texttt{exoTEDRF} and \texttt{NAMELESS} using the \texttt{ExoTEP} pipeline \citep{Benneke_2017, benneke2019a}. In total we fit six light curves, arranged as three joint fits corresponding to the three observational visits. For each visit, we simultaneously fit two light curves: the first- and second-order white light curves from each of the two NIRISS/SOSS visits and the NRS1 and NRS2 light curves from the phase curve observation captured using NIRSpec/G395H. We construct the Order 1 broadband white-light curves by summing the flux contribution across all wavelengths, while for Order 2 we restrict the wavelength coverage to 0.6-0.85\,$\mu$m \citep{Albert_2023, Radica_2023, Radica_2024}. For NIRSpec/G395H, we sum over the wavelength ranges $2.9-3.72\,\mu$m for NRS1 and $3.83-5.1\,\mu$m for NRS2 \citep{Alderson_2024}.

We fit the extracted white light curves following \citet{Radica_2024}, modeling each transit with an astrophysical and a systematics component. The transit is modeled using \texttt{batman} \citep{2015batman}, with quadratic limb-darkening coefficients fixed to values derived using \texttt{ExoTic-LD} \citep{exoticld}, the stellar parameters from \citet{Burt_2024} (i.e., $\mathrm{[Fe/H]} = 0.3, \mathrm{T_{eff}} = 3575, \mathrm{log_{10}(g_*)} = 4.85$), and the \texttt{MPS-ATLAS-1} stellar models \citep{2023mpsatlas}. Alternative fits using quadratic and non-linear power laws find no impact on the resultant transmission spectrum. We fix the orbital period to 0.66913924 days \citep{Burt_2024} and adopt a circular orbit ($e$ = 0, $w = 90^\circ $), consistent with the free-eccentricity fit of \citet{luque}. We fit the mid-transit time ($T_0$), scaled semimajor axis ($a/R_\mathrm{s}$), impact parameter ($b$), and planet-to-star radius ratio ($R_\mathrm{p}/R_\mathrm{s}$) using wide uniform priors. $T_0$, $a/R_\mathrm{s}$, and $b$ are shared across NIRISS/SOSS Orders 1 and 2 and NIRSpec/G395H detectors NRS1 and NRS2, while $R_\mathrm{p}/R_\mathrm{s}$ is fit independently for each order and detector.  

The remaining systematics component accounts for instrumental and astrophysical noise independent of the transit. For NIRISS/SOSS TSOs, we identify instrumental noise associated with JWST's 2-4 minute heating cycle and subtle drift in the trace's y-position; affected integrations are excluded as described in Section \ref{sec:exotedrf_niriss}. Although principal component de-trending suppresses short-term periodic noise, residual correlated noise persists on 10-50 minute timescales. We model this noise using a Gaussian Process (GP) with a Matérn-3/2 kernel implemented in \texttt{celerite} \citep{2017celer}, allowing for systematics and astrophysical variability, such as stellar granulation \citep{2025_ltt}. The kernel amplitude is fit independently for each order, while its characteristic timescale is shared. A linear time-dependent trend captures baseline variations, including phase variations and long-term detector ramps. Finally, an error-inflation term for each order and detector is added in quadrature to the nominal photometric uncertainties to account for unmodeled noise and provide more conservative error estimates.

We derive parameter posteriors using the \texttt{emcee} Markov Chain Monte Carlo sampler \citep{emcee}, adopting wide, uninformative priors and running 30,000 steps per joint fit, with 12,000 burn-in steps. The systematics-corrected white light curves and best-fit transit models for both visits are shown in Figure \ref{fig:lightcurves}. We combine the NIRISS/SOSS Order 1 $R_\mathrm{p}/R_\mathrm{s}$ measurements from both visits using an error-weighted mean and adopt the stellar radius $R_\mathrm{s} = 0.4555 \ R_\odot$ from \citet{Burt_2024}. The two visits agree within $<2\sigma$, yielding $R_\mathrm{p}/R_\mathrm{s} = 0.02872^{+0.00032}_{-0.00031}$ and $R_\mathrm{p} = 1.427 \pm 0.043 R_\oplus$. The inferred scaled semimajor axis and impact parameter are consistent between visits and with previous literature estimates within $1\sigma$ \citep{luque, 2026fisher}. Orbital parameter constraints from the NIRISS/SOSS and NIRSpec/G395H white light curve fits are found in Table \ref{tab:wlc_fits}.

\subsection{Spectroscopic Light Curve Fitting} \label{subsec:slc}
We generate spectroscopic light curves by binning both NIRISS/SOSS and NIRSpec/G395H data to a constant resolution of R = 100. This resolution is chosen because it achieves a high enough signal-to-noise ratio to robustly model the above-mentioned systematics; however, we note that our results remain consistent even if we bin to a higher resolution (i.e., R = 200). Finally, we note that binning-before-fitting the light curves is preferred because it supplies more accurate estimates of the transit depth and its associated uncertainties. Binning-after-fitting the light curves leads to a possible transit depth error overestimation if the covariance between data points is not properly accounted for, as it assumes individual spectral bins are independent and uncorrelated, which could be inaccurate for narrow width bins \citep{binning_madhu}. 

We fix wavelength-independent orbital parameters to their best-fit white light curve values (Table \ref{tab:wlc_fits}) and adopt bin-specific quadratic limb-darkening coefficients from \texttt{ExoTIC-LD}. Each spectroscopic light curve is jointly fit with a transit model parametrized by $R_\mathrm{p}/R_\mathrm{s}$ and a systematics model comprised of a linear time trend and a GP with a 3/2-Matérn kernel. The linear trend accounts for observed baseline variations and potential trace-position drift, particularly evident in NRS1. We fix the GP characteristic timescale to the white light curve best-fit value while allowing its amplitude to vary independently in each wavelength bin, capturing wavelength-dependent correlated noise \citep{Radica_2024,Ahrer_2025, 2025t1d, Radica_1c_2025}. We sample the fitted parameters for 10,000 steps per wavelength bin, using four walkers per parameter, and rejecting the first 5,000 steps as burn-in. Example NIRISS/SOSS spectroscopic light curves and best-fit models are shown in Figure \ref{fig:lightcurves}. Transmission spectra from the \texttt{exoTEDRF} and \texttt{NAMELESS} reductions of both NIRISS/SOSS visits are compared in Figure \ref{appendix:NL_EX_comp}. The \texttt{exoTEDRF} NIRSpec/G395H spectrum is presented in Figure \ref{appendix:NIRSPEC_TS} alongside the \texttt{Eureka!}, \texttt{Tiberius}, and \texttt{exoTIC-JEDI} reductions from \citet{luque}.

The \texttt{exoTEDRF} and \texttt{NAMELESS} transmission spectra agree closely, differing by an average of $0.65\sigma$ and $0.68\sigma$ for Visits 1 and 2, respectively. Given this agreement, which is within the $2\sigma$ uncertainties, and the slightly lower transit depth uncertainties from \texttt{exoTEDRF}, we adopt the \texttt{exoTEDRF} reduction for the remainder of our work. We construct the final NIRISS/SOSS spectrum by computing the error-weighted mean transit depth in each bin across both visits and use the \texttt{exoTEDRF} reductions of both NIRISS/SOSS and NIRSpec/G395H for the subsequent atmospheric analysis.

\section{Atmospheric \& Stellar Contamination Modeling} \label{sec:atmsec}
We use \texttt{SCARLET} and \texttt{POSEIDON} to perform one-dimensional chemically-consistent retrievals of TOI-1685b using the R = 100 NIRISS/SOSS and NIRSpec/G395H transmission spectra. We additionally investigate the effect of unocculted stellar heterogeneities by performing retrievals that assess whether the transmission spectrum can be explained by stellar activity alone or requires contributions from both the planetary atmosphere and stellar heterogeneity. 

\subsection{Atmospheric Modeling with \texttt{SCARLET}}
We utilize the atmospheric retrieval framework \texttt{SCARLET} \citep{2012benneke, 2013ApJ...778..153B, benneke2019a, 2019benb, 2021pell, 2023piaul} to explore atmospheric scenarios consistent with the NIRISS/SOSS and NIRSpec/G395H transmission spectra of TOI-1685b. We perform one-dimensional, chemically consistent equilibrium retrievals in which molecular abundances are determined by atmospheric metallicity, C/O ratio, and local temperature-pressure conditions. Our forward model solves the radiative transfer equation under hydrostatic and local thermodynamic equilibrium. We then use Bayesian parameter estimation to infer the posteriors and identify the best-fitting atmospheric model by optimizing the planetary radius at a reference pressure of 10 mbar.

We model the atmosphere using 60 layers uniformly spaced in log-pressure from $10^{-6}-10^{4}$\,bar and assume an isothermal temperature profile. The temperature prior is Gaussian, centered on the planet's equilibrium temperature of $1062$K with a standard deviation of $60$K--- T (K) $\sim \mathcal{N}$ (1062, $60^2$). We assume the background gas to be a mixture of H$_2$ and He, adopting a Jupiter-like He/H$_2$ ratio of 0.157 \citep{1996hemass} consistent with previous retrievals performed for close-in super-Earth planets \citep{toi270b, Radica_1c_2025}. Additionally we include molecules with spectral features across the NIRISS/SOSS and NIRSpec/G395H bandpass---namely H$_2$O, CO$_2$, CO, CH$_4$, NH$_3$, and SO$_2$. Molecular opacities are calculated line-by-line \citep{2013ApJ...778..153B} using \texttt{HELIOS-K} \citep{heliosk,heliosk2} cross-sections for CH$_4$ \citep{ch4}, CO$_2$ \citep{co2}, CO \citep{co}, H$_2$O \citep{polansky}, H$_2$ \citep{h2s}, HCN \citep{hcn} and NH$_3$ \citep{ammonia}, following \citet{Radica_1c_2025}. We also include collision-induced absorption from H$_2$-H$_2$ and H$_2$-He pairs due to the chosen filler gases \citep{collisions}. 

We adopt a log-uniform prior on the effective surface pressure, $\mathrm{log \ P_{eff,surf} \ (bar)} = \mathcal{U}(-8, 2)$, which defines the optically thick boundary imposed by a solid surface or clouds/aerosols. To account for potential instrumental offsets \citep{Alderson_2024, Piaulet-Ghorayeb_2024, toi270b}, we fit three additional parameters with uniform priors, $\mathrm{\delta_{O2}, \delta_{NRS1}, \delta_{NRS2}  \ (ppm)} = \mathcal{U}(-250, 250)$, representing relative offsets between NIRISS/SOSS Orders 1 and 2 and between Order 1 and the NIRSpec/G395H NRS1 and NRS2 datasets. We sample the parameter space using multiellipsoid nested sampling implemented in \texttt{nestle} \citep{2004skilling, 2006skilling}, with at least 3,000 live points. Model spectrum are initially computed at high resolution (R $=15,625$), convolved to match the resolution of the transmission spectra, and evaluated against the data using a chi-squared likelihood.

\begin{figure*}[t]
    \begin{center}
        \includegraphics[width=0.73\textwidth]{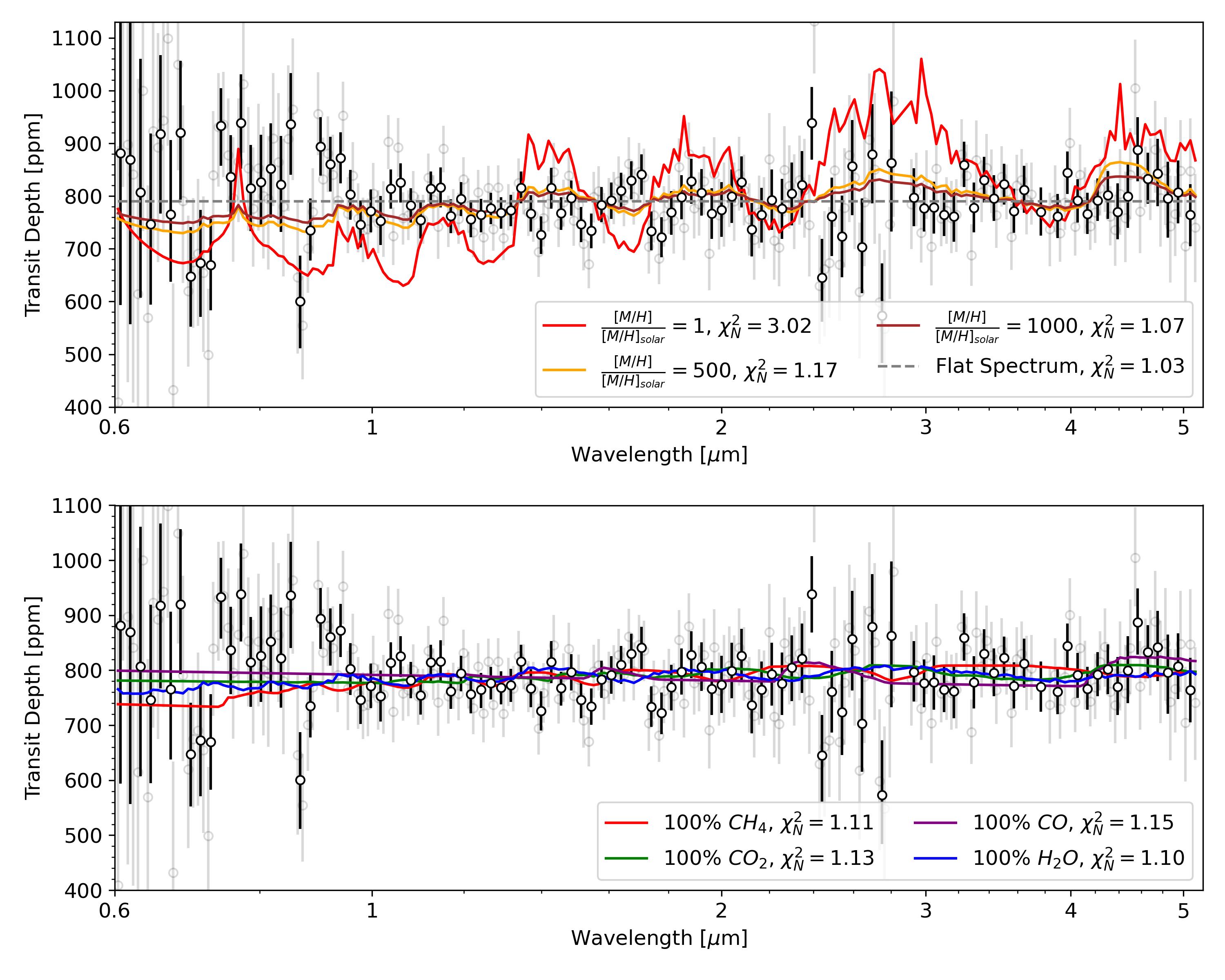}
        \caption{\textbf{Top Row:} Comparison of the \texttt{exoTEDRF} reduction combined NIRISS/NIRSpec transmission spectrum, binned to R=50 for visual clarity (black points), to cloudless \texttt{SCARLET} forward models assuming a solar carbon-to-oxygen ratio (C/O = 0.54) and atmospheric metallicity enhancements relative to solar of 1x (red), 500x (orange), and 1000x (red). The non-binned \texttt{exoTEDRF} transmission spectrum is shown in the grey, faded points. We adopt offsets of $-2 \pm 168$, $-24.4 \pm 19$, and $15.9 \pm 17$ ppm among the NIRISS/SOSS Order 1 transmission spectrum and that of Order 2, NRS1, and NRS2, respectively. The shown reduced chi-squared $\chi^2_N$ values are calculated between the models and the higher-resolution (R = 100, N = 211 points) spectra. The low amplitude spectral features are inconsistent with cloudless H$_2$/He-rich atmospheres, favoring metallicities greater than 500 times solar. \textbf{Bottom Row:} Combined NIRISS/NIRSpec transmission spectrum compared to pure 100\,bar CH$_4$ (red), CO$_2$ (green), CO (purple), and H$_2$O (blue) atmospheric forward models. The lack of identifiable spectral features does not rule-out secondary atmospheres as single-molecule models are statistically indistinguishable from the flat-line model, a scenario which shows marginal preference by the data.}
        \label{fig:Forward_Models}
    \end{center}
\end{figure*}

Prior to performing atmospheric retrievals, we compute a grid of self-consistent forward models assuming thermochemical and hydrostatic equilibrium to compare with the observed transmission spectrum (Figure \ref{fig:Forward_Models}). We consider 10 bar cloud-free atmospheres with metallicities of 1, 500, and 1000 times solar, adopting a C/O ratio of 0.54, zero Bond albedo, and a low heat redistribution efficiency ($\epsilon$ = 0.06; \citet{luque}). We use the same opacity treatment described above, including molecular absorption and H$_2$--H$_2$/H$_2$--He collision-induced absorption (CIA). We also model high MMW atmospheres dominated by a single absorber (CH$_4$, CO$_2$, CO, or H$_2$O). At T$_\mathrm{eq} \sim 1065$K, thermochemical equilibrium predicts CO and H$_2$O to dominate at pressures below $0.1$ bar, with CH$_4$ becoming the dominant C-bearing species at $\sim 0.1-1$ bar \citep{2014forward_mode_justif}. These limiting cases therefore probe plausible compositions and potential departures from equilibrium due to photochemistry or vertical mixing. 

We next investigate whether the pronounced short-wavelength slope observed in both NIRISS/SOSS visits (see Appendix Figure \ref{appendix:NL_EX_comp}) can be explained by the transit light source (TLS) effect arising from unocculted stellar heterogeneities. Cool spots and hot faculae can imprint wavelength-dependent signatures on transmission spectra, potentially mimicking or masking planetary atmospheric features \citep{2018spots, 2023RAck}. We use \texttt{stctm} \citep{piaulet_ghorayeb_stctm} to test whether stellar contamination alone can reproduce the observed spectrum, assuming either a gray, opaque cloud deck or a bare rocky surface for the planet. Details of the retrieval setup are provided in Appendix \ref{appendix:stctm_description}.

Finally, we perform joint planetary atmosphere---stellar contamination retrievals with \texttt{SCARLET} and \texttt{POSEIDON}, simultaneously fitting atmospheric and stellar heterogeneity parameters to the combined NIRISS/SOSS and NIRSpec/G395H spectrum. For \texttt{SCARLET}, we follow the approach of \citet{2023lim}, accounting for the TLS effect by applying wavelength-dependent contamination factors, $\epsilon_{\lambda, \mathrm{het}}$, calculated from the sampled parameters prior to evaluating the likelihood \citep{Piaulet-Ghorayeb_2024, Ahrer_2025}. We consider three scenarios: no stellar contamination, faculae-only contamination, and combined spot and faculae contamination. For the latter two cases, we retrieve the covering fraction, surface gravity, and temperature of each stellar heterogeneity component relative to the fitted photosphere. The \texttt{POSEIDON} retrieval setup is described in Appendix \ref{appendix:POSEIDON}.

\section{Results} \label{sec:results}

\subsection{Constraints on the Atmospheric Composition of TOI-1685b} \label{sec:atm_comp}

\begin{figure*}[t]
    \begin{center}
        \includegraphics[width=0.95\textwidth]{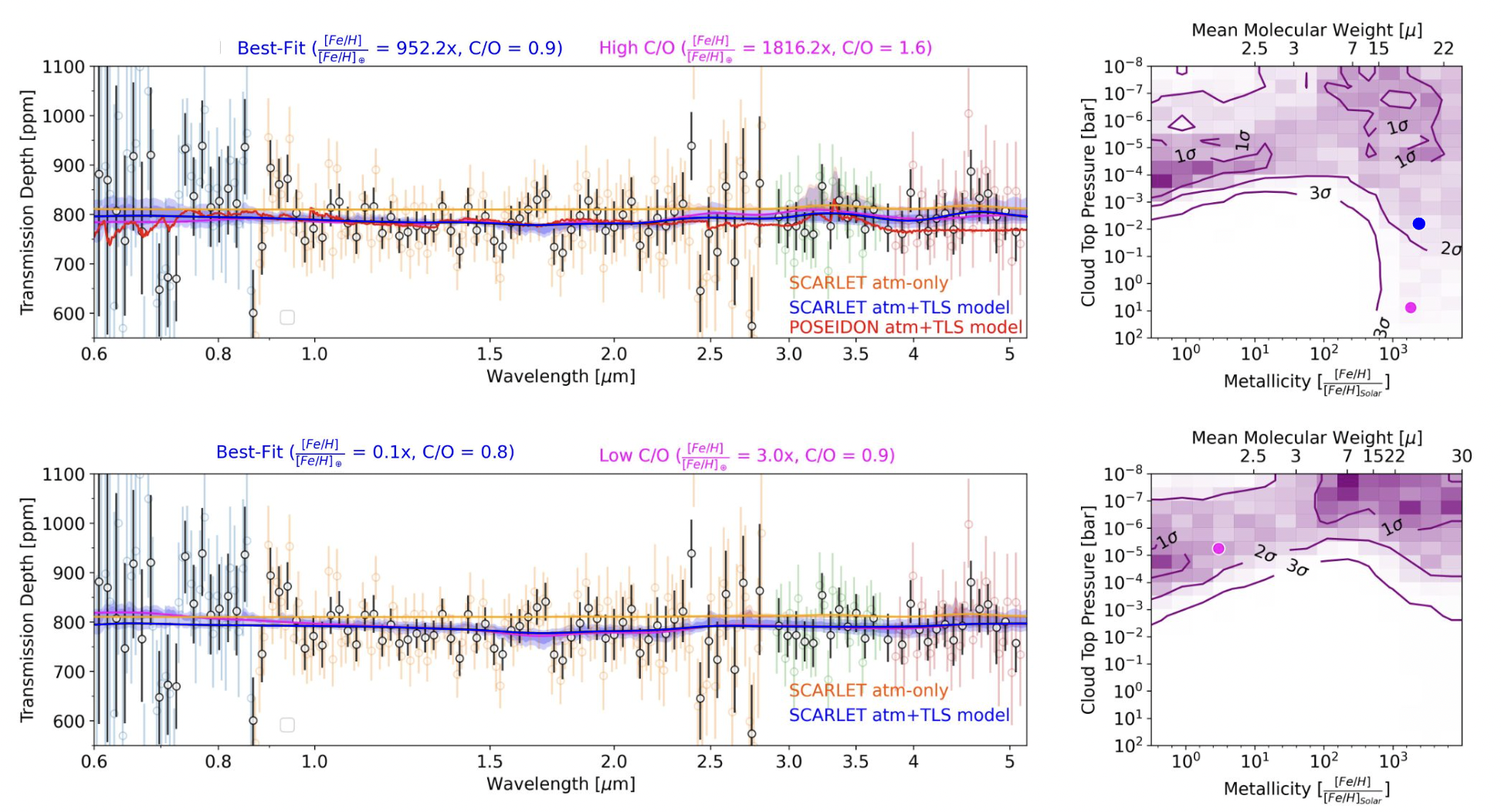}
        \caption{Chemical-equilibrium atmospheric retrievals with \texttt{SCARLET} (blue and orange) and \texttt{POSEIDON} (red) for the combined JWST NIRISS/SOSS and NIRSpec/G395H transmission spectrum of TOI-1685b. \textbf{Top Left:} Retrievals with unconstrained C/O priors, allowing exploration of high-C/O ($\geq 0.9$) solutions. Orange shows the \texttt{SCARLET} atmosphere-only model, while blue and magenta show the best-fitting joint atmosphere-stellar contamination model and an illustrative high-C/O solution, respectively. Blue shaded regions indicate the $1\sigma$ and $2\sigma$ confidence envelopes derived from posterior sample spectra. The red curve shows the best-fitting \texttt{POSEIDON} joint model. Observations are shown as points with errorbars which are colored by instrument: NIRISS/SOSS Order 2 (blue), NIRISS/SOSS Order 1 (orange), NRS1 (green), and NRS2 (red). Relative offsets from Order 1 are $-2 \pm 168$, $-24.4 \pm 19$, and $15.9 \pm 17$ ppm for Order 2, NRS1, and NRS2, respectively. Retrievals use the R=100 spectrum (colored points); models and observations are binned to R=50 (black points) for clarity. Unconstrained C/O produces more structured posteriors despite limited uniquely identifiable molecular features. \textbf{Top Right:} Joint posterior distributions of cloud-top pressure and atmospheric metallicity, with mean molecular weight indicated on the upper axis. Blue and magenta markers identify the parameter combinations corresponding to the best-fitting and illustrative models in the left panel. \textbf{Bottom Panels:} Same as above, but with C/O restricted to $< 0.9$; the magenta curve and marker now show an illustrative low-C/O solution. These retrievals favor a largely featureless spectrum and exclude cloud-free atmospheres at pressures down to 1 mbar across all metallicities at $>3\sigma$, disfavoring high-metallicity ($>540 \times$ solar; $\mu>8 \ \mathrm{g \ mol}^{-1}$), deep-cloud solutions and shifting likelihood toward reduced-scale-height atmospheres and/or high-altitude aerosols. }
        \label{fig:co}
    \end{center}
\end{figure*}

We first compare TOI-1685b's transmission spectrum with forward models, including those spanning metallicities of 1, 500, and 1000 solar, for which H$_2$O, CO, and CO$_2$ are the dominant trace absorbers, with relative abundances varying as a function of metallicity. As shown in Figure \ref{fig:Forward_Models}, the results indicate a marginal preference for a flat-line model, suggesting that the data do not robustly support the presence of detectable atmospheric features either in the form of single-species dominated atmospheres or gaseous mixtures of high metallicity. This comparison motivates the detailed chemical retrieval analysis below, where we explore the range of atmospheric compositions that can remain consistent with the largely featureless spectrum. 

We use chemically consistent equilibrium retrievals with \texttt{SCARLET} and \texttt{POSEIDON} to explore atmospheric compositions permitted by the combined transmission spectrum. The stellar C/O ratio is measured to be $0.77^{+0.40}_{-0.26}$, consistent with solar composition, providing motivation for exploring oxygen-dominated compositions while accounting for uncertainties in the host star's elemental abundances \citep{gromek2026}. Accordingly, we consider two C/O priors: $\mathrm{C/O}<0.9$, encompassing oxygen-dominated compositions consistent with near-solar elemental abundances (C/O $\approx 0.54$; \citet{2009_asplund}), and a broader prior extending to $\mathrm{C/O}=3.0$, allowing carbon-rich solutions as the C/O ratio approaches and exceeds unity \citep{2010_Carbonenh}. The resulting retrievals span substantially different metallicity and mean-molecular-weight regimes, despite producing similarly featureless transmission spectra (Figure \ref{fig:co}). In both cases, we find a strong coupling between atmospheric metallicity and C/O, such that the inferred metallicity depends sensitively on the adopted C/O prior. These results highlight the range of atmospheric compositions consistent with the data under different assumptions about the elemental inventory, which may reflect primordial protoplanetary disk chemistry or subsequent interior---atmosphere processes that alter the atmospheric C/O ratio \citep{Schlichting_2022, 2025rev, werlen2025}.

As shown in Figure \ref{fig:co}, cloud-free, H$_2$/He-dominated atmospheres with surface pressures $>1$ mbar are excluded at $>3\sigma$ for both C/O priors. The unconstrained C/O retrieval also places a lower bound of  $\mu>7.4$ g mol$^{-1}$ on the mean molecular weight (MMW). These results are consistent with previous studies of short-period, low-mass planets \citep{toi270b, Ahrer_2025, luque}. However, H$_2$-dominated atmospheres at sufficiently low surface pressures remain consistent with the data and cannot be ruled out. At 1 mbar, a H$_2$-dominated atmosphere would have an envelope mass of $\sim 7 \times 10^{15}$ kg, corresponding to a mass fraction of $\sim 4 \times 10^{-10}$. The reduced atmospheric column density could suppress molecular absorption features without invoking clouds or high metallicities. Nevertheless, such low-column-density, primordial envelopes are unlikely to survive the intense irradiation and atmospheric escape expected over the planet's lifetime \citep{2017owen,2018ginz}, even if replenished through interior outgassing \citep{2023h2loophole}. 

We caution, however, that the low-pressure solutions (Figure \ref{fig:co}, right panels) should not be taken as precise constraints on atmospheric composition. At these pressures, photochemistry, vertical mixing, and chemical kinetics may drive abundances away from equilibrium predictions, particularly in strongly irradiated atmospheres \citep{dyrekso2}. We therefore interpret these solutions primarily as evidence that a low atmospheric column density can explain the muted spectrum. Tighter constraints on the composition require retrievals that explicitly account for disequilibrium chemistry. 

Higher metallicity atmospheres remain viable within $3\sigma$ for both C/O priors. At intermediate metallicities, the retrievals favor surface pressures below 1 mbar, consistent with reduced scale heights or high-altitude aerosols that mute molecular features (Section \ref{sec:clouds}). For C/O $>1$, an additional region of solutions permits deeper ($P_\mathrm{surf}>0.1$ bar), cloud-free, high-MMW atmospheres enriched in CH$_4$ at the $3\sigma$ level. Restricting C/O to $<0.9$ largely suppresses these carbon-rich solutions, particularly above 400 times solar metallicity ($\mu \gtrsim 7$ g/mol). The retrieval, furthermore, disfavors scenarios with clouds below the observable photosphere and instead favors highly muted absorption features produced by high-altitude clouds or low-pressure, high MMW atmospheres ($P_\mathrm{surf} < 10^{-3}$ mbar). 

\subsection{Limitations on the Presence of High-Altitude Aerosols}
\label{sec:clouds}
Atmospheric metallicity influences cloud formation by regulating the abundance of condensable species, atmospheric opacity, and condensation conditions. Higher metallicity generally increases atmospheric opacity, warming the T-P profile at a given pressure and shifting condensation to lower pressures, while increasing the availability of cloud-forming material. Consequently, metallicity constraints derived from molecular absorption features (e.g., H$_2$O, CO, CO$_2$, or CH$_4$) can provide insight into the likelihood and composition of clouds. At an equilibrium temperature comparable to that of TOI-1685b ($T_\mathrm{eq} \sim 1065$\,K), plausible condensates include silicates (MgSiO$_3$, enstatite; Mg$_2$SiO$_4$, forsterite), refractory oxides (Al$_2$O$_3$, corundum), and alkali salts (Na$_2$S, ZnS, KCl, and NaCl) \citep{gao_cloud}.  

Despite the thermochemical plausibility of condensates, evidence suggests that TOI-1685b lacks an optically thick, cloud-dominated atmosphere at observable altitudes. Planetary Bond albedo generally increases with cloud coverage \citep{2010_cloud_alb, 2025_cloud_ref}; however, the low ``effective" Bond albedo inferred for TOI-1685b (A$_\mathrm{eff} = 0.2 \pm0.1$; \citealp{luque}) suggests efficient absorption of incident stellar radiation, disfavoring a highly reflective aerosol layer. This low albedo is instead consistent with either a largely cloud-free atmosphere extending to pressures P$_\mathrm{surf} \gtrsim 100$\,mbar or a  directly observable rocky surface with intrinsically low reflectivity, such as one composed of Fe-oxidized or basaltic materials \citep{luque}. 

%%% START HERE!!!!
\subsection{Influence of Stellar Contamination and Instrument Offsets} \label{sec:st_cont}

By visual inspection, TOI-1685b's transmission spectrum shows possible evidence of stellar contamination, most clearly seen by the fitted downward slope at wavelengths $< 1\mu$m (see Appendix Figure \ref{appendix:NL_EX_comp}). Unocculted faculae reduce the observed transit depth and produce stronger wavelength-dependent effects at shorter wavelengths, where emission differences arising from varying temperature gradients across the stellar surface are most pronounced. In the \texttt{stctm} retrievals, faculae reproduce the observed slope without requiring a planetary atmosphere contribution. TLS-only retrievals favor faculae over a single-temperature photosphere at wavelengths $< 1.2\mu$m, while spots are only weakly supported; accordingly, two- and three-component models yield nearly identical results (Figure \ref{appendix:tls}). However, at wavelengths $\gtrsim1.2 \mu$m in the NIRISS/SOSS bandpass, the NIRISS/SOSS spectrum is better reproduced when spots are included.

To assess the contribution of stellar contamination in the presence of a planetary atmosphere, we perform joint \texttt{SCARLET} retrievals of the NIRISS/SOSS and NIRSpec/G395H transmission spectrum. Stellar-contamination models are only weakly favored over the atmosphere-only model, with $\Delta \ln Z = 0.137$ ($1.29 \sigma$) for faculae alone and $\Delta \ln Z = 0.199$ ($1.37 \sigma$) for spots and faculae. Although these results marginally favor the inclusion of stellar heterogeneities, the evidence is insufficient to establish a significant detection of stellar contamination. Nevertheless, given their ability to reproduce features in the observed spectrum, we retain the spots and faculae scenario when interpreting the atmospheric composition in Section \ref{sec:atm_comp}. 

Our chemical-equilibrium retrievals also fit relative offsets between the NIRISS/SOSS and NIRSpec/G395H datasets to account for residual calibration differences. For the H$_2$/He-background model with spots and faculae and an unconstrained C/O prior, the retrieved offsets are $-2.2^{+167.9}_{-164.4}$, $-24.4^{+18.5}_{-15.0}$, and $15.9^{+16.6}_{-14.7}$ ppm for Order 2, NRS1, and NRS2, respectively. The offset posteriors are shown in Figure \ref{appendix:high_co_posterior} and \ref{appendix:low_co_posterior}. All offsets are consistent with zero within $2\sigma$ and are smaller than the smallest transit depth uncertainties. We therefore adopt the retrieved offsets when comparing the observations with the forward models (Figure \ref{fig:Forward_Models}) and joint atmosphere and stellar contamination models (Figure \ref{fig:co}).

\subsection{Assessment of Chi-Squared Flat-Line Significance and Residual Structure}
The forward model comparisons in Section \ref{sec:atm_comp} indicate that none of the considered atmospheric models is strongly preferred over a flat transmission spectrum. We therefore independently test whether either the \texttt{exoTEDRF} or \texttt{NAMELESS} spectrum exhibits significant wavelength-dependent structure. Specifically, we test the null hypothesis that the \texttt{exoTEDRF} and \texttt{NAMELESS} spectra are consistent with a wavelength-independent transit depth, with deviations arising from random measurement noise. We perform $\chi^2$ flat line rejection tests, together with Kolmogorov-Smirnov (K-S) and Anderson-Darling (A-D) tests on the normalized residuals \citep{1986_KS_AD}. For each error-weighted spectrum, we adopt the median transit depth as the reference flat-line model (830 ppm for \texttt{exoTEDRF} and 866 ppm for \texttt{NAMELESS}). The resulting $\chi_\nu^2$ values are 1.05 ($\chi^2=162.73$, $p=0.32$, $0.99\sigma$) and 1.02 ($\chi^2=159.07$, $p=0.39$, $0.86\sigma$), respectively, providing no significant evidence against the flat-line hypothesis. 

%S ignificant deviations from normality would instead suggest that unmodeled structure or systematic effects may contribute to the observed variance. While the normalized residuals do not, by themselves, establish that the flat-line model is correct, they provide no evidence for additional structure beyond the quoted measurement uncertainties.
Following \citet{welbanks_2025}, we report the adjusted A-D statistic to account for the relatively small number of NIRISS/SOSS data points. For \texttt{exoTEDRF}, we obtain K-S/A-D statistics of 0.04/0.21 with p-values of 0.97/0.85, while \texttt{NAMELESS} gives 0.08/0.28 with p-values of 0.26/0.66. These tests assess whether the normalized residuals are consistent with a standard Normal distribution, as expected for a flat-line model with independent Gaussian measurement errors. The low K-S statistics and high p-values provide no evidence of significant departures from normality or additional unmodeled structure beyond the quoted measurement uncertainties. Together with the $\chi^2$ tests, these results indicate that a flat spectrum cannot be rejected for either reduction.

%We also assess the possible contribution from stellar heterogeneities (such as sunspots and/or faculae) on the atmospheric retrievals. From a Bayesian model comparison, the inclusion of stellar contamination is weakly favored in our chemically consistent retrievals (Table \ref{tab:model_bayes}). Notably, the evidence against a model without stellar contamination is comparable to that obtained when evaluating a featureless (flat-line) transmission spectrum. This similarity arises, because in the absence of stellar contamination, the contribution of the planetary atmosphere becomes largely wavelength-independent and is therefore consistent with a spectrally flat signal. 

%TLS-only retrievals performed on the transmission spectrum of each visit independently yield consistent spot and faculae covering fractions, as well as corresponding temperatures (see Figure \ref{appendix:spot+fac} in the Appendix). The consistency across visits supports our assumption of a constant TLS contribution between the two NIRISS/SOSS observations. We therefore focus on the results from the TLS-only retrieval applied to the \textcolor{red}{error-propagated, mean-combined} transmission spectrum and evaluate whether stellar contamination can reproduce the observed spectral features. 

\section{Constraints on the Composition of TOI-1685b from Observed Mass and Radius} \label{sec:mr_relations}

Taken together, the lack of detectable atmospheric signatures and the low inferred albedo suggest that TOI-1685b possesses little to no substantial atmosphere. In the absence of a significant gaseous envelope or optically thick aerosols, the observed physical properties are more naturally interpreted as reflecting its intrinsic bulk composition rather than atmospheric effects. 

For a given composition, a planet's self-gravity compresses the interior, increasing the observed bulk density above the corresponding zero-pressure density. Using our NIRISS/SOSS-derived radius ($R_p=1.43\pm 0.04\,R_\oplus$) and the mass from \citet{Burt_2024} ($\mathrm{M_p}=3.03\pm0.33\mathrm{M_\oplus}$), we derive a bulk density of $\rho_P = 5.74 \pm 0.79 \ \mathrm{g/cm^3}$. However, gravitational compression at TOI-1685b's higher mass means that an Earth-like composition (32.5\% Fe and 67.5\% MgSiO$_3$ by mass) would yield a smaller radius of $1.35 \pm 0.01 \  \mathrm{R_\oplus}$ and a higher density of $\rho_{P} =6.80 \pm 0.75\ \mathrm{g/cm^3}$. Accounting for enhanced gravitational compression, the observed density corresponds to $84 \pm 14\%$ of Earth's density when scaled to $1 \mathrm{M}_\oplus$ ($\rho_{1M_\oplus} = 4.65 \pm 0.82 \ \mathrm{g/cm^3}$), indicating a modest density deficit relative to an Earth-like composition. 

We therefore explore three physically plausible interior scenarios that could reconcile the planet's observed bulk density with the apparent absence of atmospheric features: (1) a reduced iron-core mass fraction relative to Earth's, (2) the sequestration of light elements, such as hydrogen and oxygen, within the deep interior, and (3) a carbon-enriched interior, for example a soot-rich mantle or crust. Each scenario can modestly reduce the bulk density without requiring a volatile-rich atmosphere, consistent with the diverse interior compositions proposed for bare-rock planets \citep{luque}.

\subsection{Lower Mass Fractions of Metal Cores} \label{sec:cmf}
Using the mass measurement of TOI-1685b reported by \citet{Burt_2024}, we construct a suite of interior structure models to predict the planet's radius across a range of solid compositions. These models explore varying initial rock and silicate abundances motivated by the chemical environment of the host star. We adopt a simplified two-layer interior structure model following the methodology of \citet{Schlichting_2022}, parameterizing the bulk composition in terms of the iron core mass fraction (CMF), defined as the ratio of iron mass to the planet's total mass. 

For the planetary iron core, we adopt the Vinet equation of state (EOS) to model the behavior of liquid Fe, using a zero-pressure reference density of $\rho_0 = 7.2 \pm 0.1$ g cm$^{-3}$, based on experimental results for molten Fe at high pressures and temperatures \citep{Schlichting_2022}. The Vinet EOS \citep{vinet1987, Vinet_1989, 2007ApJ...669.1279S} is given by  $$P = 3K_0\eta^{2/3}(1 - \eta^{-1/3}) \times \exp\left[\frac{3}{2} (K_0' - 1)(1 - \eta^{-1/3})\right],$$ where $\eta = \rho/\rho_0$ is the compression ratio, $K_0$ is the isothermal bulk modulus, and $K_0'$ is its pressure derivative. The Fe-rich core is modeled using an EOS derived from experimental measurements of molten Fe at high pressures and temperatures, rather than the high-pressure solid Fe phase, as this treatment more accurately reproduces the mass-radius relationship for terrestrial planets. For the silicate mantle, we employ the third-order Birch-Murnaghan EOS to model the behavior of MgSiO$_3$ perovskite \citep{birch1947, 2007ApJ...669.1279S}. The Birch-Murnaghan EOS is expressed as $$P = \frac{3}{2}K_0(\eta^{7/3}-\eta^{5/3})\left[1+\frac{3}{4}(K_0'-4)(\eta^{2/3}-1)\right].$$ In both layers, $\eta$ is defined relative to the corresponding uncompressed density. We neglect any Fe-Mg partitioning (i.e. variations in Fe/(Fe + Mg)) within the silicate layer and assume a pure perovskite mantle. 
%The derivation of both EOS formulations is detailed in \citet{Poirier_2000}, and the experimental data from which these models are constrained primarily span pressure regimes up to 150 GPa and temperatures below 2000 K - conditions that are broadly consistent with the pressure-temperature regimes expected within the modeled planets. Together, this dual-layer EOS framework enables a physically grounded and self-consistent mapping of the mass-radius relationship for rocky planets. 

For a given EOS, we integrate the above equations from the center of the planet using the boundary conditions of $m(0) = 0$ and $P(0) = P_c$, where $P_c$ is the central pressure. The outer boundary condition is $P(R_\mathrm{p}) = 0$. The planet's final mass $M_\mathrm{p}$ and radius $R_\mathrm{p}$ are uniquely determined by a specific central pressure (i.e. $P(0) = P_c$) and metal core mass fraction. We omit any temperature dependence in our EOS, and strictly separate the metal and silicate contents between two layers (i.e. planetary interior is fully differentiated). 

\begin{figure*}[ht]
    \begin{center}
        \includegraphics[width=0.85\textwidth]{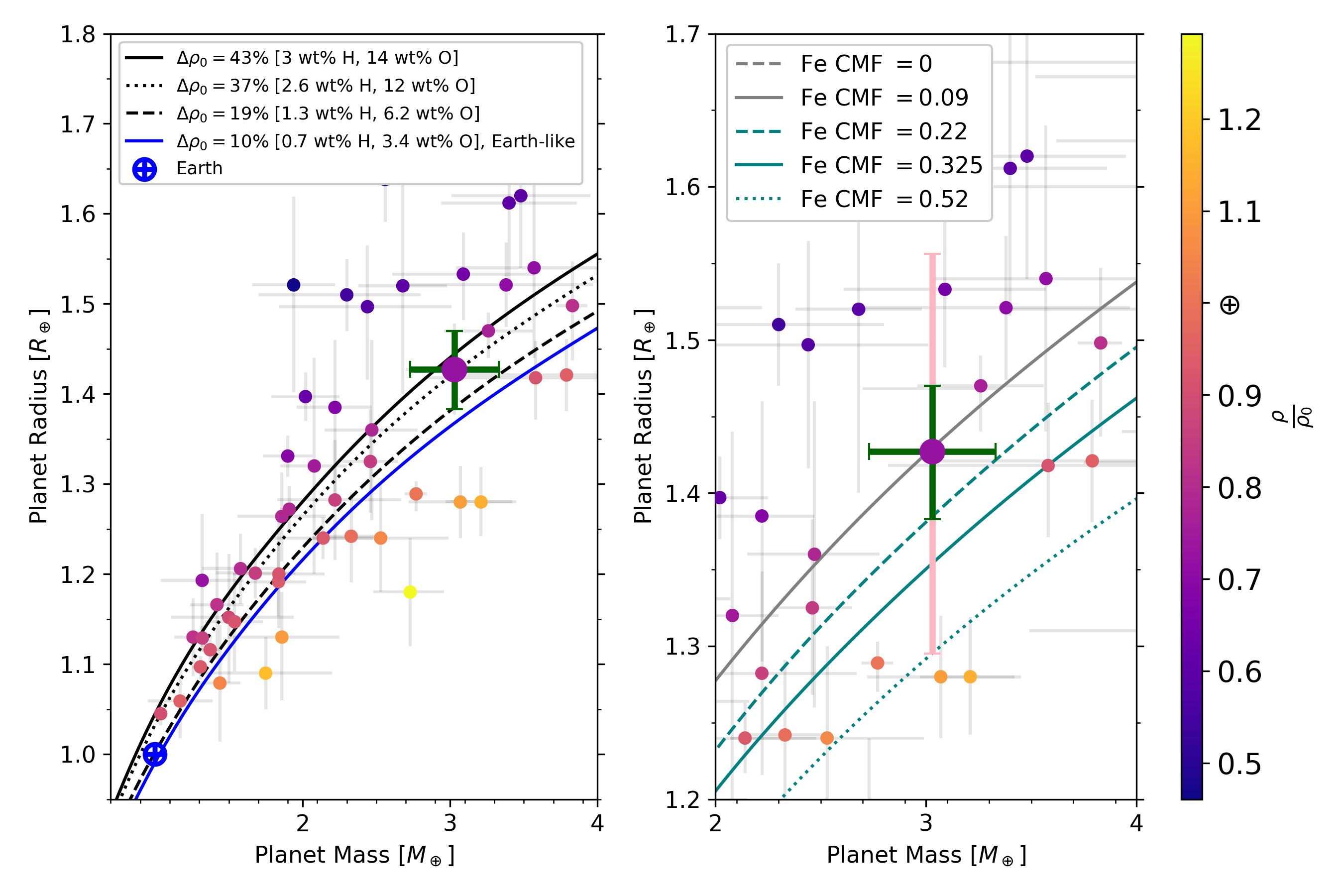}
        \caption{Mass-radius relationships for differentiated bare rock planets. \textbf{(Left Panel)} Curves show mass-radius relations for planets with Earth-like core mass fractions ($32.5\%$), but varying core density deficits ($10-43 \%$) resulting from the incorporation of H and O into the metallic core (corresponding to $0.7-3$ wt$\%$ H and $3-14$ wt$\%$ O by mass). The solid blue line represents an Earth-like composition with a solid Fe core and MgSiO$_3$ perovskite mantle. The dashed, dotted, and solid black lines indicate progressively larger hydrogen and oxygen contents in the metal, producing increasing density deficits in the Fe core parametrized by $\Delta \rho_0$. Earth is shown for reference. \textbf{(Right Panel)} Curves show the mass-radius relations for planets with pure Fe core mass fractions (CMF) ranging from 0 to 0.52. The derived $1\sigma$ ($3\sigma$) radius constraints of TOI-1685b are shown in green (pink) and are consistent with reduced Fe CMFs of $\leq$ 0.22 ($\leq 0.52$). For an Earth-like CMF, TOI-1685b's $1\sigma$ constraints are compatible with core density deficits of 19-43 $\%$, corresponding to 1.3-3 $\%$ H and 6.2-14$\%$ O. The high metallicity of the host star ([Fe/H] $=0.3 \pm 0.1$) appears in tension with solutions requiring reduced iron core mass fractions and therefore favors models with core density deficits due to light elements into the core \citep{fleeting_not_forgotten}. Also shown are exoplanets from the NASA Exoplanet Archive with mass and radius uncertainties $\leq 30 \%$ and $\leq 10 \%$, respectively. Point colors indicate each planet’s nominal bulk density relative to Earth’s using $M_p/M_{\oplus} = (R_p/R_{\oplus})^4(\rho_p/\rho_\oplus)^{4/3}$ \citep{VALENCIA2006545}. }
        \label{fig:differentiated}
    \end{center}
\end{figure*}

Our interior structure models suggest that TOI-1685b is consistent with CMFs ranging from 0 to 0.22 within the 1$\sigma$ uncertainty and from 0 to 0.52 within the 3$\sigma$ uncertainty of the measured radius (Figure \ref{fig:differentiated}). Furthermore, the nominal range of plausible CMFs is generally lower than Earth's core mass fraction ($\mathrm{CMF_\oplus} \approx 0.325 \pm 0.003$; \citealt{2018_earth_cmf}). In the absence of a significant atmosphere, the measured mass and radius therefore favor a silicate-rich, iron-poor interior. However, even a modest gaseous envelope would measurably increase the apparent transit radius. For plausible atmospheric temperatures and high-$\mu$ atmospheric compositions, a 1-mbar atmosphere would extend $\sim50-90$km ($\sim 0.005-0.009$ R$_p$), while a 1-bar atmosphere itself contributes of order $\sim20$km ($\sim 0.002$ R$_p$) to the measured radius. Such radial contributions are comparable to the current radius uncertainties and would permit a broader range of CMFs---including iron-rich interiors---to remain compatible with observations. 

Despite the formal consistency of low-CMF models with the observed radius, such compositions are disfavored given the metal-rich nature of the host star ([Fe/H] $=0.3 \pm 0.1$; \citet{Burt_2024}). The iron content of rocky planets is thought to correlate with the bulk metallicity of their protoplanetary disks, which in turn is reflected by the host star's metallicity [Fe/H]. Planets forming in metal-rich environments are expected to accrete from solids with higher iron abundances, therefore higher iron core mass fractions are expected \citep{2023met_cmf}. In contrast, metal-poor systems appear to preferentially host planets consistent with low iron core masses, suggesting an upper limit on CMFs imposed by disk composition \citep{fe_cmf, 2026ApJ_1to1}. Therefore, the low-CMF solutions for TOI-1685b, though permissible based on mass-radius constraints alone, are disfavored given the host star's high metallicity, particularly in the absence of additional mechanisms such as mantle stripping.

\subsection{Sequestration of Light Elements in the Metal Core}
The low bulk densities observed in some super-Earths have traditionally been attributed to either reduced core mass fractions (see Section \ref{sec:cmf}) or to variations in silicate composition arising from spatially-dependent condensation in the protoplanetary disk \citep{dorn}. However, \citet{Schlichting_2022} demonstrated that such planets may instead retain Earth-like CMFs if their iron cores have sequestered substantial amounts of light elements, particularly hydrogen and oxygen. Chemical equilibration between a planet's hydrogen-rich atmosphere, silicate melt, and metallic core enables hydrogen to dissolve into the melt and partition into the metal, while redox reactions associated with water formation drive the concurrent incorporation of oxygen, establishing a steady-state exchange among the envelope, mantle, and core \citep{Schlichting_2022}. If hydrogen is efficiently transported from the envelope to the deep interior, significant hydrogen and oxygen inventories can accumulate in the core. In this scenario, a core density deficit of order a few percent to tens of percent is plausible if the planets initially accreted hydrogen-rich envelopes comprising a few percent of their total mass, which were subsequently lost through atmospheric escape \citep{2017phot,2018ginz,2019gup, 2020gup,2021phot}. 

The resulting reduction in core density can substantially alter the inferred mass-radius relationships of such planets, complicating interpretations that rely solely on bulk composition. To investigate the observational consequences of hydrogen and oxygen assimilation into a super-Earth's iron core, we construct a suite of interior models similar to those described in Section \ref{sec:cmf}. Experimental and theoretical constraints indicate that the incorporation of light elements, such as hydrogen and oxygen, into iron significantly reduces the uncompressed density of the metal. In particular, \citet{li2019} found that the addition of hydrogen decreases the uncompressed density, $\rho_0$, by approximately 8.7$\%$ per wt$\%$ H in Fe metal, while oxygen reduces $\rho_0$ by about $1.2\%$ per wt$\%$ O. Following the approach of \citet{Schlichting_2022}, we parametrize the core density deficit as a fractional reduction of the initial uncompressed density $\Delta \rho_0$, within a two-layer differentiated planet model. 

While thermodynamics may allow for large light-element fractions in the metallic core, a physical upper limit is imposed by the requirement that the core density remain equal to or greater than that of the mantle at the boundary of the two-layer differentiated interior model. We adopt the same maximum core density deficit as \citet{Schlichting_2022}, $\Delta\rho_0 = 43\%$. The mass-radius relationship corresponding to this maximum core density deficit, assuming an Earth-like CMF, is shown as the black solid line in the left panel of Figure \ref{fig:differentiated}. This limiting case corresponds to a core composition containing 3$\%$ hydrogen and $14\%$ oxygen by mass in the metallic core. Consistent with \citet{Schlichting_2022}, we find that this mass-radius relationship closely matches that of a planet composed entirely of MgSiO$_3$ perovskite. 

Figure \ref{fig:differentiated} shows that the measured bulk density of TOI-1685b can be reproduced by models that incorporate hydrogen and oxygen into the metal core. Assuming an Earth-like core mass fraction ($\mathrm{CMF} = 0.325$), TOI-1685b's observed radius is consistent within 1$\sigma$ uncertainties with iron core density deficits ranging from 19-43\%. This range corresponds to approximately 1.3-3 wt$\%$ hydrogen and 6-14 wt$\%$ oxygen incorporated into the metallic core. In particular, a moderate density deficit of $\sim37\%$---equivalent to about 2.6 wt$\%$ H and 12 wt$\%$ O by mass in the metal core---is sufficient to reproduce TOI-1685b's observed bulk density. 

Nevertheless, the current mass and radius measurements do not uniquely constrain the planet's interior structure. As discussed in Section 6.1, the $1\sigma$ radius uncertainty is consistent with CMFs up to 0.22, whereas the $3\sigma$ uncertainty permits CMFs extending beyond Earth-like. The density deficits quoted above are contingent upon the adopted Earth-like CMF: lower CMFs require smaller density deficits to reproduce the observed radius, while higher CMFs require larger density deficits. The high metallicity of the host star disfavors lower CMFs solutions and makes compositions requiring a core-density deficit more likely. 

To provide context, we overplot exoplanets from the NASA Exoplanet Archive\footnote{\url{https://exoplanetarchive.ipac.caltech.edu/}} with well-constrained mass and radius measurements ($\leq 30\%$ and $\leq 10\%$ uncertainties, respectively). The data points are color-coded by the bulk density of the planet relative to Earth's, calculated using the relationship defined in \citet{VALENCIA2006545}: $\rho_p/\rho_{\oplus} = (M_\mathrm{p}/M_{\oplus})^{3/4}(R_\mathrm{p}/R_{\oplus})^{-3}$. This comparison highlights the substantial diversity in bulk densities among rocky planets---including TOI-1685b---and illustrates how variations in core density deficit can reproduce observed mass-radius trends. However, we note that the inferred interior properties are contingent upon the adopted radius measurement. As highlighted by \citet{2024edwards}, derived radii may be affected by instrument-dependent systematics, and therefore the interior models presented here should interpreted within the context of the current observational uncertainties. To illustrate the sensitivity of our conclusions to the adopted radius, we compare our radius measurement with recent literature values and present the implications for the inferred interior structure in Appendix \ref{appendix:radius_comparison}.

Recent work on carbon-rich planet formation beyond the refractory organic carbon ``soot line" and snow line, together with observations of carbon-enhanced protoplanetary disks around M dwarfs with elevated gas-phase C/O ratios ($\geq 0.54$; \citealp{Bergin_2023, 2023tabone, refId0}), suggest that carbon-rich interiors may also account for intermediate bulk densities  \citep{SeagerSoot25}. We evaluate TOI-1685b as a potential soot-rich super-Earth by comparing its mass and radius with the soot-rich mass-radius (M-R) relations of \citet{soot25arxiv}. Comparing these models with the measured properties of TOI-1685b shows that the planet is denser than even the most extreme diamond-like soot scenarios. We therefore conclude that a carbon-rich, soot-dominated interior is unlikely to explain TOI-1685b's observed bulk density. 

\section{Conclusion} \label{sec:conclusion}
We present the first JWST/NIRISS transmission spectrum of TOI-1685b and find it to be largely featureless, consistent with either a high-metallicity atmosphere,  high-altitude aerosols, or a barren rocky surface. Cloud-free, hydrogen-rich atmospheres are robustly excluded at $>3\sigma$ confidence for pressures down to 1\,mbar, reinforcing predictions from atmospheric escape models that TOI-1685b should no longer retain a primordial envelope. However, cloud-free secondary atmospheres---such as those dominated by H$_2$O, CO$_2$, or CH$_4$---remain viable. The observational degeneracy between muted molecular features and optically thick clouds continues to complicate the interpretation of flat transmission spectra for short-period rocky exoplanets. 

However, the low effective Bond albedo of TOI-1685b (A$_\mathrm{eff} = 0.2 \pm0.1$; \citealp{luque}) suggests that the planet is not highly reflective and may lack a bright cloud deck, increasing the likelihood that we are probing the planet's surface directly. High-opacity condensates such as MgSiO$_3$, Al$_2$O$_3$, or alkali salts could plausibly form aerosols that mute spectral features under intermediate to high-metallicity conditions, but the current emission data undoubtedly favor the classification of the planet as a barren rocky world over this scenario. Additionally, the hot dayside ($0.99 \pm 0.07$ times that of a perfect blackbody; \citet{luque}) of TOI-1685b further supports the planet's lack of a primordial atmosphere and existence as a bare rock.

We explore three plausible bare-rock scenarios capable of reproducing TOI-1685b's observed bulk density without requiring a substantial atmosphere: (1) a reduced iron core mass fraction, (2) sequestration of light elements (such as hydrogen and oxygen) into the interior, and (3) a carbon-enriched interior. The first two scenarios are consistent with the measured mass and radius, whereas the carbon-rich scenario produces densities inconsistent with observations. We favor the incorporation of lighter elements into the metal core over a reduced iron core mass fraction because the host star's high metallicity ([Fe/H] = $0.3 \pm 0.1$) disfavors formation of a planet with a substantially depleted iron inventory. 

Models of coupled atmospheric escape and interior evolution demonstrate that the loss of a primordial H$_2$-rich envelope can drive progressive oxidation and facilitate core-mantle-atmosphere interactions, allowing material to be exchanged between otherwise chemically distinct reservoirs \citep{fleeting_not_forgotten}. If such reactions are thermochemically favorable, a fraction of the initially volatile-rich material in the atmosphere can ultimately be sequestered into the interior even as the atmosphere is depleted. This provides a pathway for producing the modest density deficit for TOI-1685b without requiring a present-day atmosphere or formation beyond the ice line. Given the system's estimated age of $1 \pm 0.3$ Gyr \citep{Burt_2024}, complete or near-complete removal of a primordial atmosphere through photoevaporation and/or core-powered mass loss is consistent with the expected timescales for these processes \citep{2024Burn, 2024venturini}. TOI-1685b could have therefore formed near its present-day location and evolved to its apparent bare rock composition.

More broadly, our results highlight an important limitation of atmospheric characterization as a probe of rocky planet composition and formation. Although atmospheres provide the most direct observable window into rocky planets, their present-day compositions need not preserve a unique record of either their primordial bulk volatile inventory or the planet's formation environment. Atmospheric escape, volatile sequestration into the interior, and phase-dependent chemical exchange between the core, mantle, and atmosphere can decouple present-day observable atmospheric abundances from a planet's initial volatile inventory and drive atmospheric evolution. Consequently, atmospheric non-detections alone cannot rule out volatile-rich formation pathways, but may instead reflect post-formation processing and sequestration of volatiles into the interior.

The decoupling also presents an observational challenge. The higher MMW secondary atmospheres expected for many close-in rocky planets are intrinsically challenging to detect in transmission, and extensive JWST observations may be required to confidently detect or rule out such atmospheres. Complementary observations in thermal emission with instruments such as JWST/MIRI could provide constraints on surface mineralogy and geology, although these measurements may require a substantial number of visits for planets such as TOI-1685b \citep{luque}. Together, these challenges underscore the need for a combined approach that links atmospheric observations with self-consistent atmosphere-interior models to reconstruct the formation and evolutionary pathways of short-period rocky planets orbiting M-dwarfs.

%% Please use the acknowledgment and contribution environments. This will 
%% be anonomyized when the "anonymous" style option is used. 
\begin{acknowledgments}
This work is based on observations with the NASA/ESA/CSA James Webb Space Telescope, obtained at the Space Telescope Science Institute (STScI) operated by AURA, Inc. All of the data presented in this paper were obtained from the Mikulski Archive for Space Telescopes (MAST). The specific observations analyzed can be accessed via \dataset[doi:10.17909/wc0a-6c49]{https://doi.org/10.17909/wc0a-6c49}. Data reductions and resultant models available upon request. H.M.A. and H.E.S gratefully acknowledge support from JWST-GO-04098.005. Y.M. acknowledges support from the European Research Council (ERC) under the European Union’s Horizon 2020 research and innovation programme (grant agreement no. 101088557, N-GINE). C.P.-G. acknowledges support from the E. Margaret Burbidge Prize Postdoctoral Fellowship from the Brinson Foundation. L.-P.C. acknowledges support from Mitacs through the Mitacs Accelerate program, in partnership with the Montreal Planetarium.

\end{acknowledgments}

\facilities{JWST}

\software{ \texttt{astropy} \citep{2013A&A...558A..33A}, \texttt{celerite} \citep{celerite}, \texttt{exoTEDRF} \citep{Radica2024a}, \texttt{matplotlib} \citep{2007CSE.....9...90H}, \texttt{numpy} \citep{2020numpy}, \texttt{pymultinest} \citep{2016pymulti}, \texttt{scipy} \citep{2020scipy}
          }


\clearpage
\appendix
\renewcommand{\thefigure}{A\arabic{figure}}
\setcounter{figure}{0}
\renewcommand{\thetable}{A\arabic{table}}
\setcounter{table}{0}

\section{NIRISS/SOSS \texttt{NAMELESS} Reduction}
\label{appendix:NAMELESS_NIRISS}

\begin{figure*}[h]
    \begin{center}
        \includegraphics[width = 1\textwidth]{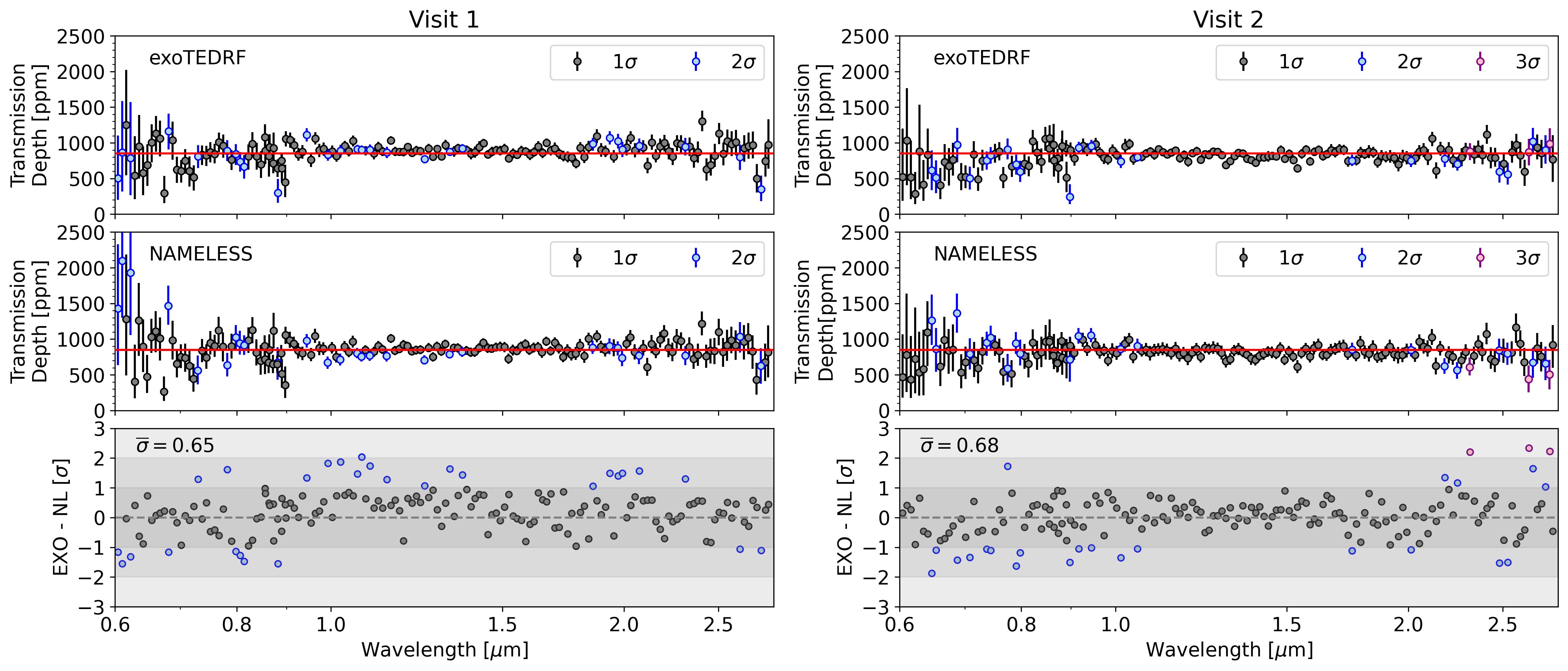}
        \caption{Comparison between the transmission spectra extracted from the \texttt{exoTEDRF} and \texttt{NAMELESS} reductions. \textbf{\textit{First row:}} \texttt{exoTEDRF} transmission spectrum across Visits 1 and 2 colored by the deviation from the \texttt{NAMELESS} spectrum. Grey points illustrate spectra that are consistent within $1\sigma$, blue points represent bins that are consistent within $2\sigma$, and pink points represent consistency within $3\sigma$. The sigma deviation is calculated using the maximum error bar between the \texttt{exoTEDRF} and \texttt{NAMELESS} reduction for each wavelength bin. \textbf{\textit{Middle row:}} \texttt{NAMELESS} transmission spectrum across Visits 1 and 2 colored by the deviation from the \texttt{exoTEDRF} spectrum.For both top and middle rows, a horizontal line at 850 ppm is included as a visual reference for comparison with the error-weighted mean transmission spectrum shown in Figure \ref{fig:Forward_Models}. \textbf{\textit{Bottom row:}} Difference in each wavelength bin divided by the maximum error bar between the \texttt{exoTEDRF} and \texttt{NAMELESS} reductions. Not only are all spectral bins within $3\sigma$, but there exist no visible systematic biases in the residual differences, signifying good agreement between the two transmission spectra reduced with independent pipelines. The average deviation between the two spectra is 0.65 and 0.68 for Visit 1 and Visit 2, respectively.} 
        \label{appendix:NL_EX_comp} 
    \end{center}
\end{figure*}

\begin{figure*}[t]
    \begin{center}
        \includegraphics[width = 0.7\textwidth]{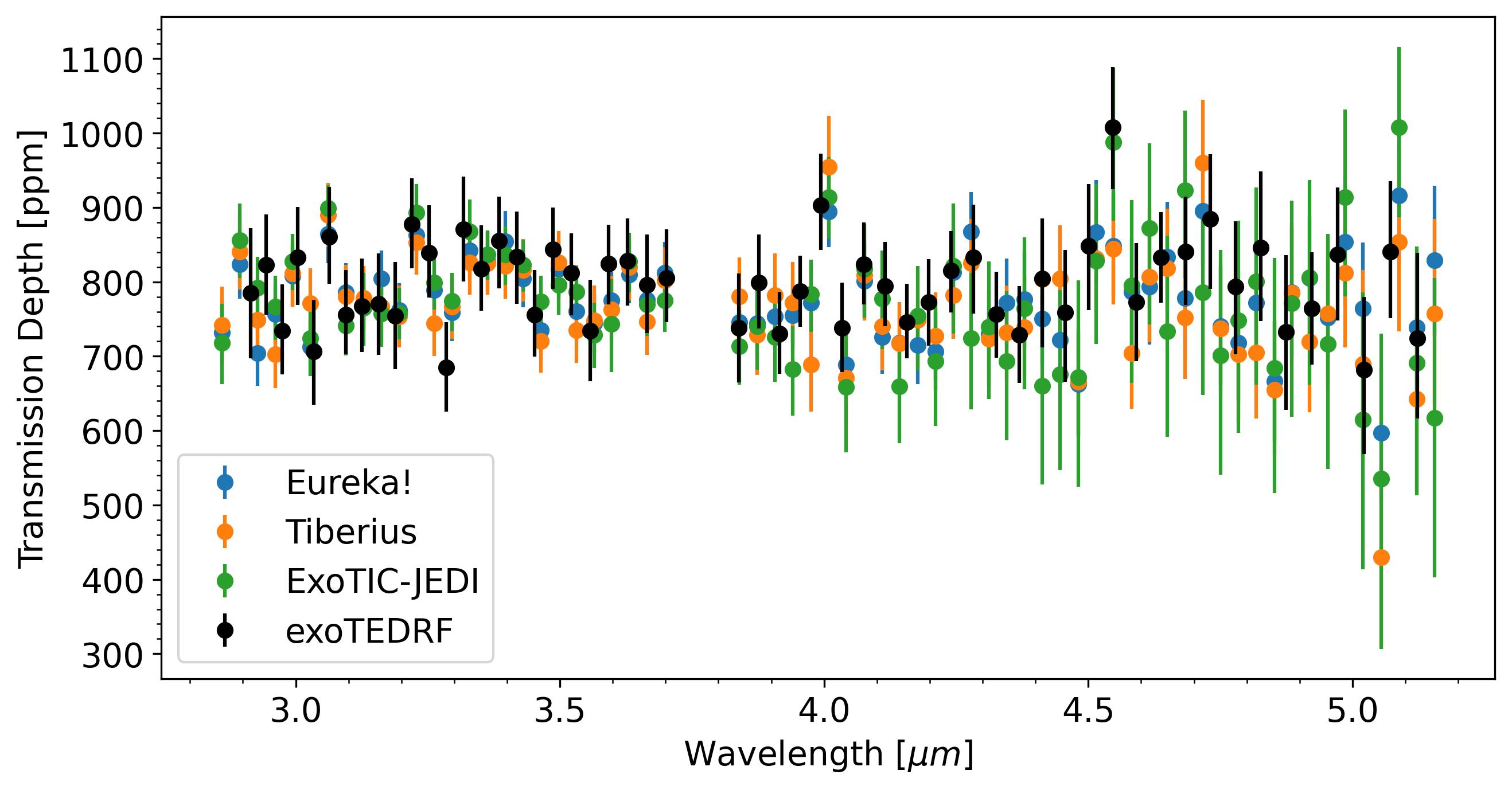}
        \caption{Comparison of the JWST NIRSpec/G395H transmission spectra extracted using four independent reduction pipelines: \texttt{exoTEDRF}, \texttt{Tiberius}, \texttt{Eureka!}, and \texttt{exoTIC-JEDI}. For the \texttt{exoTEDRF} reduction, we apply the NRS1-2 offset of $\approx 40.3$ppm derived from the joint stellar contamination and atmospheric retrieval with \texttt{SCARLET}. The transmission spectra for the \texttt{Tiberius}, \texttt{Eureka!}, and \texttt{exoTIC-JEDI} reductions are taken from \citet{luque}.  All four transmission spectra are consistent within $1\sigma$, and the residuals show no evidence of systematic biases, indicating strong agreement across the independent reductions.}
        \label{appendix:NIRSPEC_TS}
    \end{center}
\end{figure*}

\begin{figure*}[h]
    \begin{center}
        \includegraphics[width=0.78\textwidth]{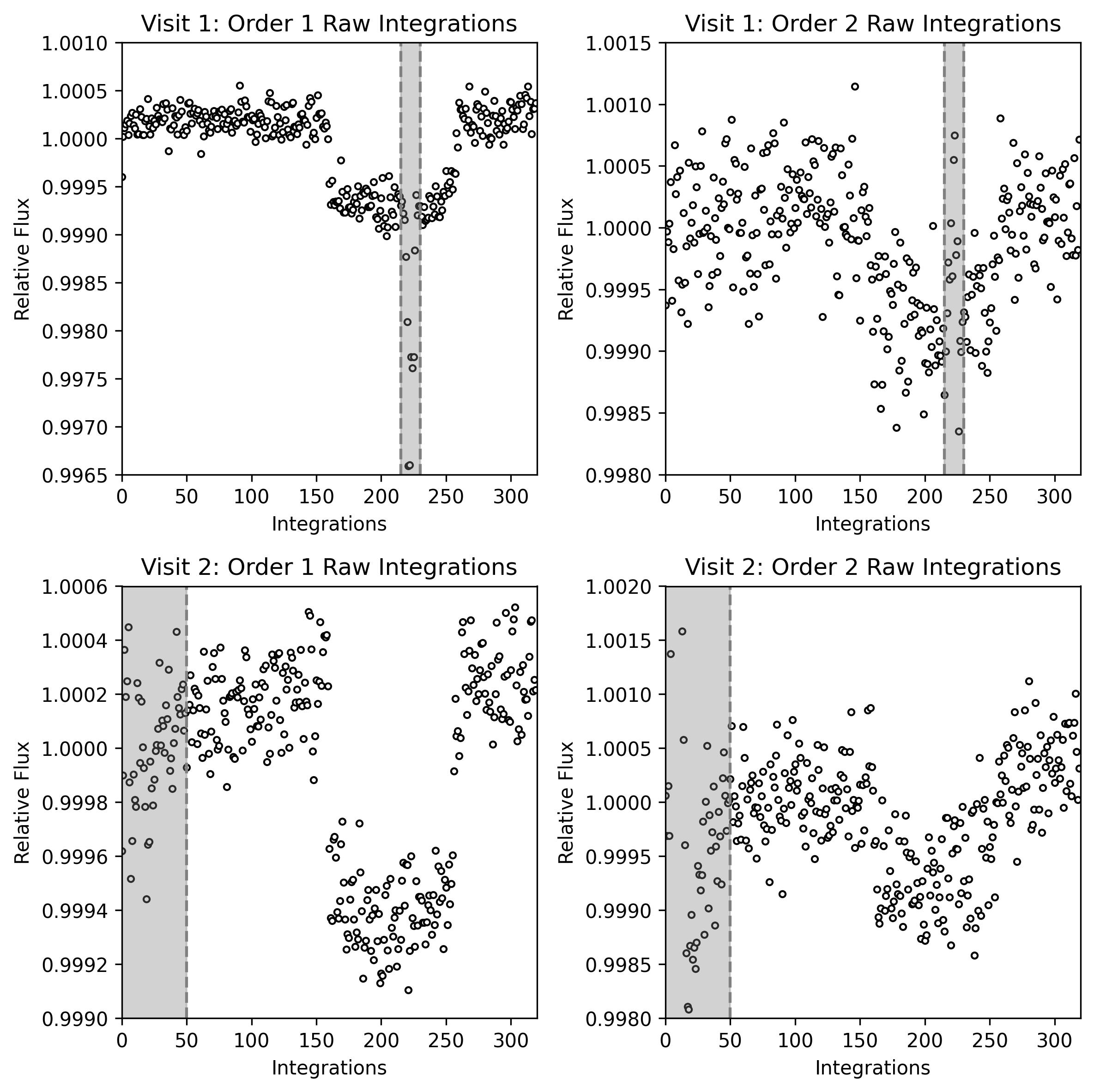}
        \caption{Raw white light curves for Order 1 and Order 2 from each visit taken with JWST NIRISS/SOSS. Grey-shaded regions indicate integrations that are affected by irreparable systematics and will be masked and excluded from further light-curve fitting using \texttt{EXOTEP}, as discussed in Section \ref{sec:lcs}. 
}
        \label{appendix:raw_integ}
    \end{center}
\end{figure*}

We perform an independent data reduction of the two NIRISS SOSS visits of TOI-1685\,b using the \texttt{NAMELESS} pipeline, closely following the methodology outlined in \citet{Coulombe2023} and \citet{2025_ltt}. We first proceed through the stage 1 and 2 steps of the \texttt{jwst} pipeline, up to the flatfield correction step. We then flag bad/hot pixels by isolating pixels with consistent outlier 2D spatial second derivatives throughout the TSOs, and correct them via bicubic interpolation. The background is subtracted from all integrations using the model from STScI, which was scaled independently to the two regions of the detector that are separated by the sharp jump in background flux around column $x\sim700$. Remaining cosmic rays are corrected by computing the running median in time of all individual pixels, and bringing all $>4\sigma$ outliers to the median's value. We then correct for 1/$f$ noise by scaling all columns of the order 1 and 2 traces individually, and fitting for an additive constant via $\chi^2$ minimization that is then subtracted from the frames. Finally, we extract the 2D stellar spectra using a fixed boxed aperture of width 38 pixels, and compute the wavelength solution using the \texttt{PASTASOSS} package \citep{baines2023characterizationvisittovisitstabilitygr700xd}. Both the Visit 1 and Visit 2 transmission spectrum, as reduced with \texttt{NAMELESS} and \texttt{exoTEDRF}, as well as the differences between visits, are shown in Figure \ref{appendix:NL_EX_comp}. The NIRSpec/G395H reductions are shown in Figure \ref{appendix:NIRSPEC_TS}. The \texttt{exoTEDRF}-reduced NIRISS/SOSS broadband light curves are shown in Figure \ref{appendix:raw_integ}, with the masked regions shaded in gray. The fitted orbital parameters from joint fits to both NIRISS/SOSS orders and the transit-only NIRSpec/G395H observations are listed in Table \ref{tab:wlc_fits}.

%%%%% NIRSPEC/G395H Transmission Spectra

%%% JOINT NIRISS/SOSS + NIRSPEC Params
\begin{table*}[h]
\centering
\caption{Fitted Orbital Parameters for TOI-1685b derived from JWST NIRSpec/G395H (NRS1 and NRS2) transit-only phase curve observations and NIRISS/SOSS (Orders 1 and 2) transit observations. Results are based on data reduced with \texttt{exoTEDRF} and modeled using \texttt{EXOTEP}.}
\begin{tabular}{l c c c } 
 \hline\hline
Fitted Parameter & Visit 1 (NIRISS/SOSS) & Visit 2 (NIRISS/SOSS) & Transit-only (NIRSpec/G395H)  \\
 \hline
Mid-transit Time $T_c$ ($BJD_{TDB}$) &  $2460691.15342^{+0.00023}_{-0.00017}$ & $2460691.82222^{+0.00026}_{-0.00024}$ & $2460355.9147^{+0.00010}_{-0.0043}$ \\ [0.5ex]
Planet-to-star radius ratio $R_p / R_*$: & & & \\  
--NIRISS/SOSS Order 1 & $0.02918^{+0.00050}_{-0.00046}$ & $0.02826^{+0.00041}_{-0.00041}$ & - \\
--NIRISS/SOSS Order 2 & $0.0294^{+0.0012}_{-0.0010}$ &  $0.02821^{+0.00082}_{-0.00076}$ & -  \\
--NIRSpec/G395H NRS1 & - & - & $0.02857_{-0.00060}^{+0.00053}$ \\
--NIRSpec/G395H NRS2 & - & - & $0.02789_{-0.00052}^{+0.00049}$  \\
  Semimajor axis ratio $a/R_*$ & $5.12^{+0.50}_{-0.88}$& $5.42^{+0.29}_{-0.51}$ & $5.63^{+0.13}_{-0.32}$ \\
  Impact parameter b & $0.44^{+0.19}_{-0.23}$ & $0.37^{+0.17}_{-0.17}$ & $0.23^{+0.17}_{-0.16}$ \\
 \hline\hline

\end{tabular}
\label{tab:wlc_fits}
\end{table*}

\section{Comparison of Published Radius Measurements for TOI-1685b}
\label{appendix:radius_comparison}
\begin{figure}[h]
    \begin{center}
        \includegraphics[width = 0.37\textwidth]{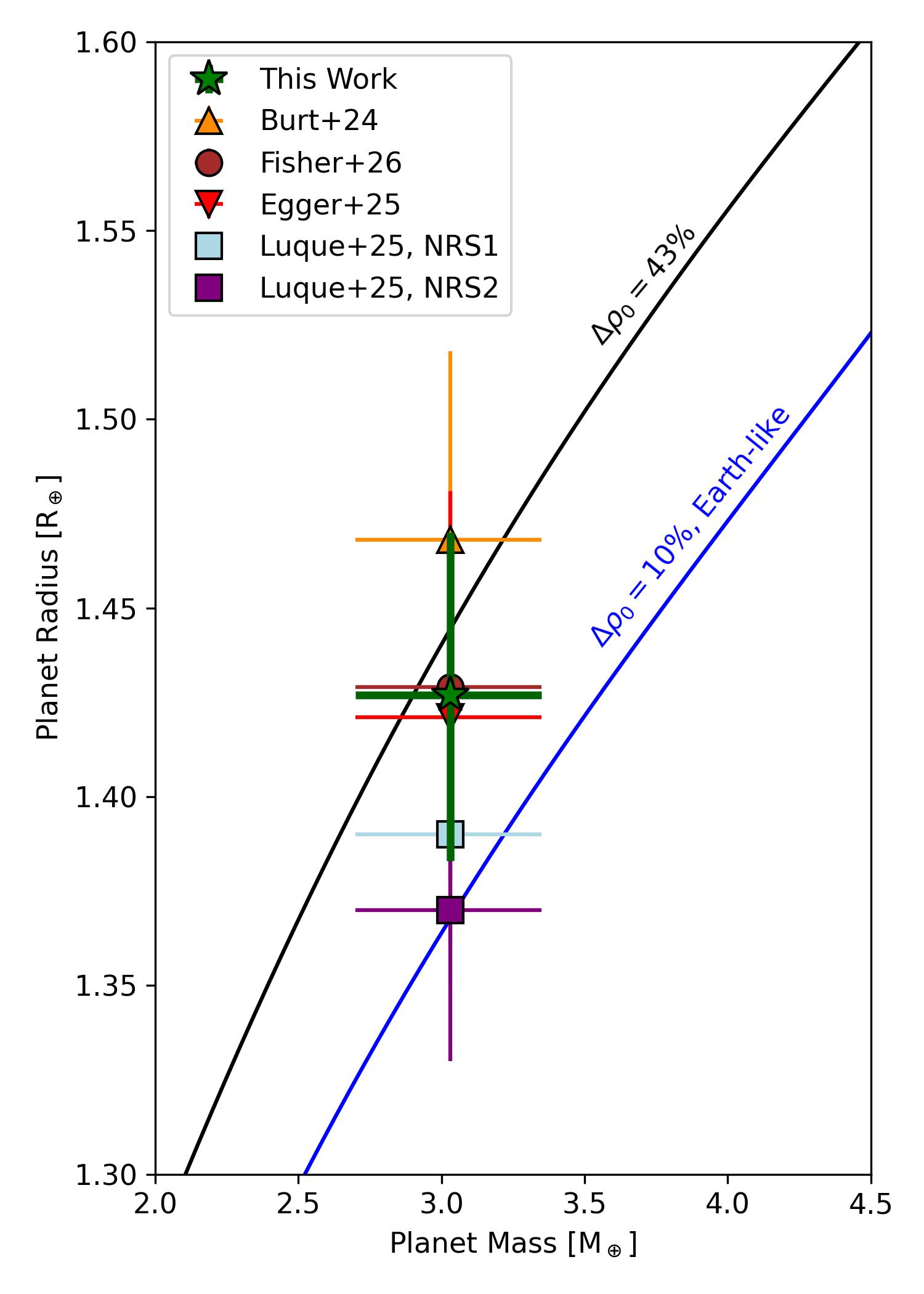}
        \caption{Comparison of published radius measurements for TOI-1685b from this work and literature (Burt+24 \citep{Burt_2024}, Fisher+26 \citep{2026fisher}, Egger+25 \citep{2025egger}, and Luque+25 \citep{luque}), assuming the same planet mass ($3.03^{+0.33}_{-0.32} \ \mathrm{M}_\oplus$; \citet{Burt_2024}). The mass-radius curves for core density deficits of $43\%$ and $10\%$ (Earth-like) are plotted in black and blue, respectively. The spread in radii illustrates how observational uncertainties propagate onto the inferred interior composition. }
        \label{appendix:rad_comp}
    \end{center}
\end{figure}

The inferred interior structure of TOI-1685b depends directly on the adopted planetary radius. As discussed by \citet{2024edwards}, transmission-derived radii may be susceptible to instrument-dependent systematics that, if uncorrected, can introduce wavelength-dependent biases in the inferred planetary radius. To place the interior modeling presented in Section 6 into this context, Figure \ref{appendix:rad_comp} compares our radius measurement with recent determinations reported by \citet{Burt_2024, 2025egger, luque, 2026fisher}, all of which adopt the mass reported by \citet{Burt_2024}. The published radius estimates span a modest range, resulting in different inferred bulk densities and consequently different preferred interior models. Our radius measurement lies above the Earth-like $10\%$ core density-deficit relation, motivating lower core mass fractions and/or the incorporation of light elements into the core presented in Section 6. In contrast, the smaller radius reported by \citet{luque} intersects the Earth-like $10\%$ core density-deficit relation, consistent with their conclusion that TOI-1685b is compatible with an Earth-like bulk density.

\section{Stellar-Contamination Only Retrievals with \texttt{stctm}} \label{appendix:stctm_description}
To assess the impact of the Transit Light Source Effect (TLSE) on TOI-1685b's transmission spectrum, we perform stellar-contamination-only retrievals on the visit-combined NIRISS/SOSS transmission spectrum using the \texttt{stctm} module \citep{piaulet_ghorayeb_stctm}. The NIRISS/SOSS transit observations and the NIRSpec/G395H phase-curve measurement were obtained nearly one year apart--- significantly longer than the stellar rotation period. This separation, together with $\sim 20$ppm offsets between the NIRISS and NIRSpec spectra, suggests that the spot and faculae coverage likely differed between the two dates. Because NIRSpec lacks the short-wavelength coverage required to robustly constrain TLS parameters when combined with NIRISS/SOSS data, we exclude the NIRSpec transmission spectrum from the stellar-contamination-only retrievals, similar to \citet{Ahrer_2025}. To test whether stellar heterogeneites alone can account for the deviations from a flat-line spectrum, we run three stellar-contamination-only retrievals assuming (1) the presence of only unocculted sunspots, (2) only faculae, and (3) both components contributing significantly. In each case, we follow the stellar surface parameterization outlined in Appendix C of \citet{Piaulet-Ghorayeb_2024}.

\begin{figure*}[t]
    \begin{center}
        \includegraphics[width=0.91\textwidth]{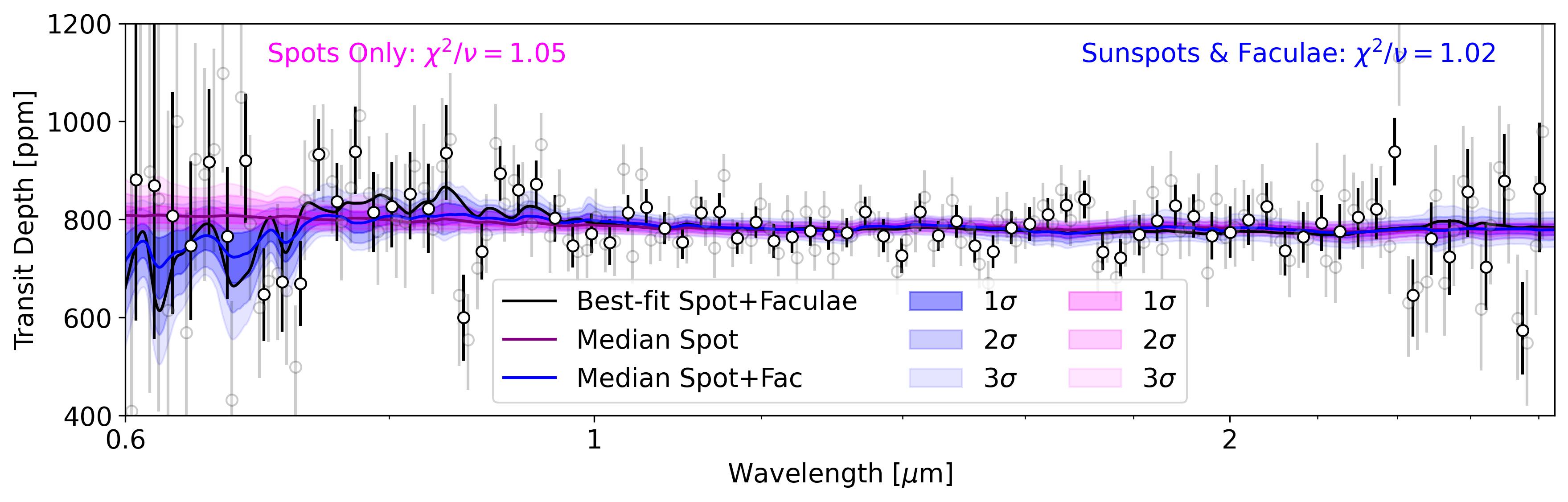}
        \caption{\texttt{stctm} retrieval results from TLS-only fits to the combined NIRISS/SOSS transmission spectrum of TOI-1685b. The panel shows both the spot-only retrieval and the joint spots+faculae retrieval. We show the best-fitting model spectrum (black line), median (purple/blue line), and the 1$\sigma$, 2$\sigma$, and 3$\sigma$ constraints derived from the posterior sample spectra (semitransparent shading) for each retrieval result, respectively. The retrievals are performed on the R = 100 data (gray shaded points), but are binned for visual clarity to R = 50 (black points). No offset is applied between orders 1 and 2. The reduced $\chi^2$ value for the spot-only (spot+faculae) retrieval are indicated in the upper left (upper right) of the panel, with $\nu$ denoting the degrees of freedom (spot-only: $\nu=150$, faculae+spots: $\nu=148$). In both cases, the NIRISS/SOSS spectrum of TOI-1685b is well reproduced by stellar contamination alone, with faculae in particular accounting for the downward slope at wavelengths $<1\mu$m. }
        \label{appendix:tls}
    \end{center}
\end{figure*}

In the faculae-only case, we fit for 6 parameters: faculae covering fraction $f_{fac}$, faculae temperature relative to the photosphere $\Delta T_\mathrm{fac} >0$, wavelength-independent transit depth D (i.e., zero-point or opaque-body reference), local gravity of faculae relative to the surface $\Delta \log(g_\mathrm{fac})$, photospheric temperature $T_\mathrm{phot}$, and local surface gravity $\log(g)$. In the combined case (i.e. spots \& faculae), we fit for two additional parameters, spot temperature relative to photosphere $\Delta T_\mathrm{spot} <0$ and local gravity of spot relative to surface $\Delta \log(g_\mathrm{spot})$. For optimal computational efficiency, we created a grid of interpolated \texttt{PHOENIX} stellar models at a metallicity 
\clearpage
[Fe/H] = 0.3 and local surface gravity log(g) = 4.85 \citep{bluhm_2021} using the \texttt{MSG} module \citep{Townsend2023}. These precomputed models span, in 20 K increments, a range of effective temperatures $T_\mathrm{eff} \in [2300, 4375]$\,K representative of TOI-1685's true photospheric properties. To generate the synthetic stellar contamination spectrum, we select the stellar model within the grid whose $T_\mathrm{eff}$ is closest in magnitude to the sampled $T_\mathrm{phot}$ and/or $T_\mathrm{fac}$. We show the $1\sigma$, $2\sigma$, and $3\sigma$ confidence regions derived from sample spectra representative of the posterior distribution of the inhomogeneity in various shades of blue in Figure \ref{appendix:tls}. 

We explore the parameter space using the \texttt{emcee} module \citep{emcee} with Markov Chain Monte Carlo (MCMC) sampling using 20 times as many walkers as there are fitted parameters. The chains are run for 7000 steps, with the first 60\% regarded as burn-in. We use a uniform prior in linear space for the fractional coverage and temperature of each heterogeneity, adopting the same constraints on $|\Delta T_\mathrm{het}| >100$ as \citet{Piaulet-Ghorayeb_2024} to guarantee differentiation from photospheric properties. Since TLS retrievals are not sensitive to the absolute temperature of the host star, $T_\mathrm{phot}$, but rather the temperature contrast between different parts of the star, we enforce a gaussian prior using the median effective temperature ($T_\mathrm{eff} = 3575$\,K; \citet{Burt_2024}) with a wider standard deviation ($\sigma = 60$\,K) analogous in approach to \citet{2023lim,  Radica_1c_2025}. 

\section{Complete Posterior Distributions for Joint \texttt{SCARLET} Retrievals}
\begin{figure*}[h]
    \begin{center}
        \includegraphics[width = 1.0\textwidth]{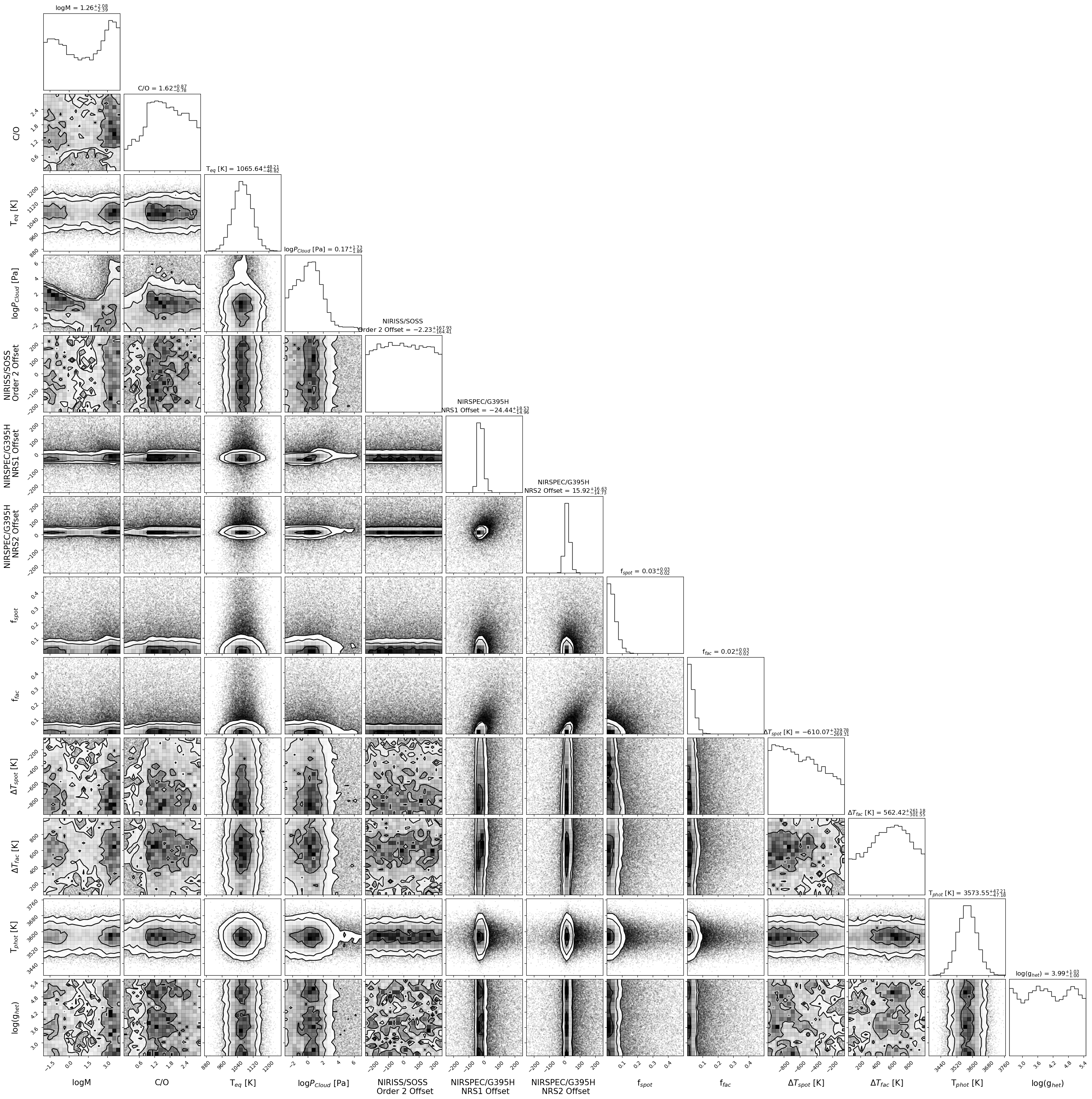}
        \caption{Posterior distributions from the joint planetary atmosphere and stellar contamination retrieval with \texttt{SCARLET} described in Section \ref{sec:atmsec}, with the C/O ratio prior left unconstrained up to a value of C/O $=3.0$. }
        \label{appendix:high_co_posterior}
    \end{center}
\end{figure*}

\begin{figure*}[h]
    \begin{center}
        \includegraphics[width = 1\textwidth]{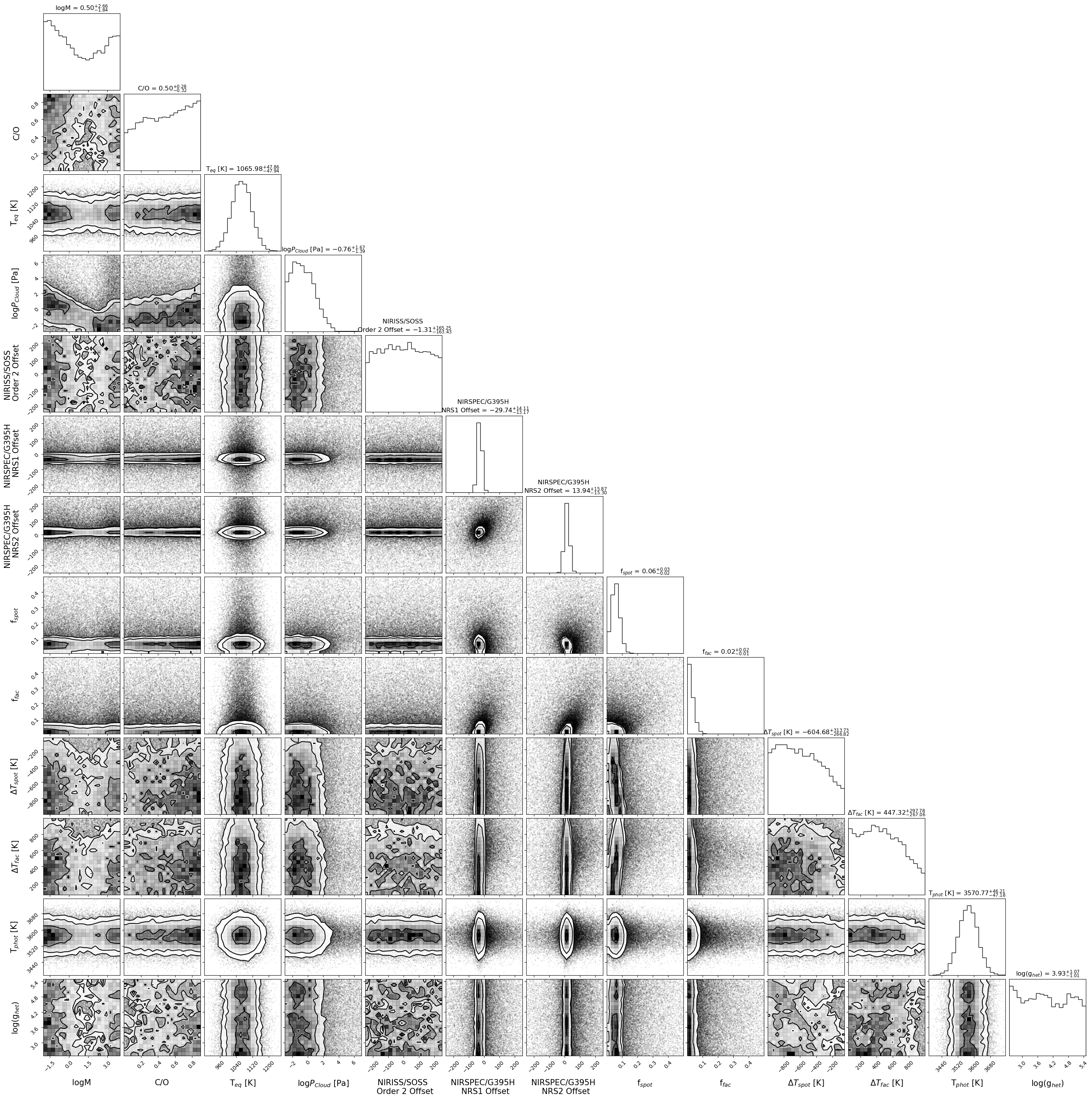}
        \caption{Posterior distributions from the joint planetary atmosphere and stellar contamination retrieval with \texttt{SCARLET} described in Section \ref{sec:atmsec}, with the C/O ratio constrained to C/O $< 0.9$. }
        \label{appendix:low_co_posterior}
    \end{center}
\end{figure*}

Figures \ref{appendix:high_co_posterior} and \ref{appendix:low_co_posterior} present the full two-dimensional and marginal posterior distributions from the joint planetary atmosphere and stellar contamination \texttt{SCARLET} retrievals described in Section \ref{sec:atmsec}. The two retrievals differ only in their assumed C/O priors. Figure \ref{appendix:high_co_posterior} shows the results obtained with an unconstrained C/O prior, whereas Figure \ref{appendix:low_co_posterior} shows the corresponding posteriors when the C/O ratio is restricted to values below $0.9$. 

\section{\texttt{POSEIDON} Atmospheric Retrieval Description and Comparison}
\label{appendix:POSEIDON}
We also conduct independent atmospheric retrievals using the open-source retrieval code \texttt{POSEIDON} \citep{MacDonald_2017, MacDonald_2023}. For each retrieval, we employ 2000 live points with \texttt{PyMultiNest}, a nested sampling algorithm that facilitates both parameter estimation and model comparison for our purposes \citep{Feroz_2009, Buchner_2014}. A more detailed explanation of the radiative transfer method and forward model (\texttt{TRIDENT}) used in \texttt{POSEIDON}, along with the associated opacity database, is provided in \citet{MacDonald_2022}.

Using the \texttt{exoTEDRF} reductions of both the NIRISS/SOSS and NIRSpec/G395H datasets outlined in Sections \ref{sec:exotedrf_niriss} \& \ref{sec:exotedrf_nirspec}, we generate model spectra across 0.6 to 5.3\,$\mu$m at a wavelength grid resolution of R = 20,000 for opacity sampling, following previous POSEIDON analyses (e.g., \citealt{Coulombe2023,Piaulet-Ghorayeb_2024,Ahrer_2025}). These model spectra are then convolved with a Gaussian kernel to match the native resolution of each instrument and scaled by the instrument sensitivity function, allowing us to bin the spectra at the desired wavelength spacing. We include all applicable collision-induced absorption opacity pairs that are available in POSEIDON \citep{MacDonald_2022,2019Karman}.

In our retrievals of TOI-1685\,b, we account for a broad range of atmospheric and stellar parameters. In each retrieval, we fit the reference radius (R$_\mathrm{p,ref}$) at a reference pressure of 10 mbar. In order to compare atmospheric and non-atmospheric interpretations of the data, we perform both multi-gas atmospheric retrievals and retrievals with the planetary atmosphere disabled on the NIRISS/SOSS and NIRSpec/G395H datasets. Our multi-gas atmospheric models include H$_2$, He, H$_2$O, CH$_4$, CO$_2$, SO$_2$, CO, and NH$_3$, with H$_2$ and He treated as background gases at a fixed He/H$_2$ ratio of 0.17\, which fills the remaining atmospheric abundance after accounting for all the retrieved molecular species. For the POSEIDON v1.3 opacity database used here, we adopt the line list from \citet{polansky} for H$_2$O, \citet{2024Yurchenko} for CH$_4$, \citet{co2} for CO$_2$, \citet{2016Underwood} for SO$_2$, \citet{Li2015} for CO, and \citet{ammonia} for NH$_3$.

Additionally, each model with an atmosphere includes both clouds and a two-parameter (log a$_\mathrm{haze}$ and $\gamma$) power-law prescription for hazes \citep{MacDonald_2017}. For the cloud layers, we assume an optically-thick gray opacity, setting all cloud layers deeper than P$_\mathrm{surf}$ to infinite opacity. For the planetary atmosphere, we assume an isothermal temperature profile. Lastly, we allow for three relative offsets ($\delta_{rel}$) between NIRISS/SOSS Order 2 and Order 1, between Order 1 and NIRSpec/G395H NRS1, and between NRS1 and NRS2, respectively, as additional free parameters.

The priors for our models of TOI-1685\,b are as follows: R$_{\mathrm{p,ref}}$ (R$_\oplus$) = $\mathcal{U}$ (0.88, 1.76), T (K) = $\mathcal{N}$ (1062, 60), log a$_\mathrm{haze}$ = $\mathcal{U}$ (-4, 8), $\gamma_\mathrm{haze}$ = $\mathcal{U}$ (-20, 2), log P$_\mathrm{cloud}$ (bar) = $\mathcal{U}$ (-7, 2), $\delta_{rel}$ (ppm) = $\mathcal{U}$ (-200, 200), and log X = $\mathcal{U}$ (-12, 0), the last of which corresponds to a broad molecular mixing ratio prior that allows each species to become the dominant atmospheric constituent. We note that a Gaussian prior is used for the atmospheric temperature of the planet, as initial retrievals were underestimating the equilibrium temperature of the planet by several hundred kelvin.

We also perform joint atmosphere and stellar contamination retrievals to assess the impact of the TLS effect on the NIRISS transmission spectrum. As all of our retrievals are jointly run on both the NIRISS/SOSS and NIRSpec/G395H datasets, we include stellar contamination in every retrieval (except in nested retrievals to compute the Bayesian significance of the stellar heterogeneities). The \texttt{POSEIDON} model accounts for stellar contamination by multiplying the bare-rock transmission term, ($R_\mathrm{p}$/$R_\mathrm{s}$)$^2$, by the wavelength-dependent stellar contamination factor from stellar spots and faculae. The model spectra of the active stellar regions are generated through \texttt{POSEIDON}'s utilization of the \texttt{PyMSG} package, which interpolates across the PHOENIX grid of stellar atmosphere models from \citet{Husser_2013}. We assume a fixed stellar surface gravity of log g = 4.85 $\pm$ 0.04 (cgs) \citep{bluhm_2021}, which is used when interpolating the PHOENIX stellar atmosphere spectra.

Consequently, we introduce five additional free parameters, using a T$_*$ = 3575 $\pm$ 53 \citep{Burt_2024}: f$\mathrm{_{fac}}$ = $\mathcal{U}$ (0.0, 0.5), f$\mathrm{_{spot}}$ = $\mathcal{U}$ (0.0, 0.5), T$\mathrm{_{fac}}$ (K) = $\mathcal{U}$ (T$_*$-3T$_{*,err}$, 1.2T$_*$), T$\mathrm{_{spot}}$ (K) = $\mathcal{U}$ (2300, T$_*$+3T$_{*,err}$), and T$\mathrm{_{phot}}$ (K) = $\mathcal{U}$ (T$_*$-T$_{*,err}$, T$_*$+T$_{*,err}$). This results in a total of 17 free parameters included in our retrievals of TOI-1685\,b. The best-fit model for \texttt{POSEIDON} is shown in Figure \ref{fig:co}.

%% For this sample we use BibTeX plus aasjournalv7.bst to generate the
\clearpage 
\bibliography{sample701}
\bibliographystyle{aasjournal} %% Check suppress initials 



\end{document}